\documentclass[12pt,preprint]{aastex}

\usepackage{bm}
\usepackage[caption=false]{subfig}
\usepackage{amsmath}

\shorttitle{Prediction of Solar Flare Size and Time-to-Flare}
\shortauthors{Al-Ghraibah et al.}

\begin{document}


\title{Prediction of Solar Flare Size and Time-to-Flare Using Support Vector Machine Regression}


\author{Laura E. Boucheron and Amani Al-Ghraibah}
\affil{Klipsch School of Electrical and Computer Engineering, New Mexico State University, Las Cruces, NM 88003}

\and

\author{R. T. James McAteer}
\affil{Department of Astronomy, New Mexico State University, Las Cruces, NM 88003}
%
%




\begin{abstract}
We study the prediction of solar flare size and time-to-flare using 38 features describing magnetic complexity of the photospheric magnetic field.  This work uses support vector regression to formulate a mapping from the 38-dimensional feature space to a continuous-valued label vector representing flare size or time-to-flare.  When we consider flaring regions only, we find an average error in estimating flare size of approximately half a \emph{geostationary operational environmental satellite} (\emph{GOES}) class.  When we additionally consider non-flaring regions, we find an increased average error of approximately 3/4 a \emph{GOES} class.  We also consider thresholding the regressed flare size for the experiment containing both flaring and non-flaring regions and find a true positive rate of 0.69 and a true negative rate of 0.86 for flare prediction.  The results for both of these size regression experiments are consistent across a wide range of predictive time windows, indicating that the magnetic complexity features may be persistent in appearance long before flare activity.  This is supported by our larger error rates of some 40 hr in the time-to-flare regression problem.  The 38 magnetic complexity features considered here appear to have discriminative potential for flare size, but their persistence in time makes them less discriminative for the time-to-flare problem.  
\end{abstract}




\section{Introduction}
\label{intro}
Solar flares are powerful events resulting from the sudden conversion of stored magnetic energy to particle acceleration and excess radiation in the corona, and releasing up to $10^{32}$ total erg of energy~\citep{fletcher2011}.  Solar flares can have significant detrimental effects on earth, including disruption and damage to satellites, power grids, and telecommunications infrastructure~\citep{space_weather_catalog,schrijver2014}.  Since coronal magnetic field measurements are not currently available, much research has focused on the inference of coronal magnetic field structure from the photospheric magnetic field.  In this, it is assumed that turbulent motions in the photosphere create complex photospheric magnetic configurations~\citep{mcateer2010}, and this will then be reflected in complex coronal magnetic configurations~\citep{mcateer2015a}.  Complex coronal field configurations store the energy that is released as a solar flare.  

\subsection{Prediction of Flare Size Properties}
Several researchers have studied the use of Poisson statistics to relate flare rate of different active region (AR) classes (e.g., McIntosh or Mount Wilson classes) to probability of observing a C-, M-, or X-class flare~\citep{wheatland2000,bloomfield2012}.  Other researchers have studied the relationship between quantitative image parameters and the probability of observing a C-, M-, or X-class flare~\citep{qahwaji2007,colak2009,song2009,mason2010}. Several researchers have also studied the relation between quantitative image parameters and flare size~\citep{sammis2000,mcateer2005,georgoulis2007,schrijver2007,welsch2009} or flare index~\citep{guo2006,jing2006}. To the best of our knowledge, no researchers have studied the direct \emph{prediction} of flare size on the basis of quantitative image parameters as we study here.

\subsection{Prediction of Flare Time Properties}
Several researchers have performed a qualitative study of the time evolution of flare parameters by plotting parameters versus time~\citep{hagyard1999,sammis2000,abramenko2003,lekabarnes2003a,meunier2004,abramenko2005,barnes2006,guo2006,wang2006,schrijver2007,conlon2008,hewett2008,jing2010,mason2010}.  Other researchers have studied flare prediction in light of different predictive time windows~\citep{barnes2006,colak2009,al-ghraibah2015}.  Related to flare time properties, several researchers have studied the relation between quantitative image parameters and flare productivity~\citep{,mcateer2005,cui2006,mcateer2015b} and flare rates~\citep{wheatland2000,wheatland2005,falconer2011}.  To the best of our knowledge, no researchers have studied the direct prediction of time-to-flare on the basis of quantitative image parameters as we study here.

In this paper, we use support vector machine regression (SVR) to predict solar flare size and time-to-flare from quantitative measures of magnetogram complexity. The remainder of this paper is organized as follows.  In Section \ref{sec:SVM}, we briefly introduce regression in general, SVR in particular, and other applications of SVR to time series analysis. In Section \ref{sec:data_experiment}, we explain our data set, experimental procedures, and error measures. In Section \ref{sec:results} we present our results of solar flare size prediction and solar flare time regression.  In Section \ref{sec:conclusions}, we discuss our conclusions and future work.

\section{Support Vector Regression (SVR)}
\label{sec:SVM}
\subsection{Regression}
In the general form of regression, we wish to predict some continuous target variable $t$ given some $D$-dimensional input vector $\mathbf{x}$~\citep{bishop2006}.  In this work, we wish to predict the flare size (``size regression'') or time to flare occurrence (``time regression'') given a set of features describing magnetic complexity.  The equation $y(\mathbf{x})$, which maps input data to the target variable, is determined through optimization of some criterion based on training data.

Regression commonly takes the general form~\citep{bishop2006}
\begin{equation}
 y(\mathbf{x},\mathbf{w})=\sum_{j=0}^{M-1}w_j\phi_j(\mathbf{w})=\mathbf{w}^\top\bm{\phi}(\mathbf{x}),
\end{equation}
where $y$ is a function whose output indicates the target value $t$ for a given input vector $\mathbf{x}=[x_1,x_2,\ldots,x_D]^\top$; $\mathbf{w}=[w_0,w_1,\ldots,w_{M-1}]^\top$ is a vector of weights applied to the basis functions $\phi_j$; basis functions $\phi_j$ are some (linear or nonlinear) function of input data $\mathbf{x}$; and $\bm{\phi}=[\phi_0,\phi_1,\ldots,\phi_{M-1}]^\top$.  In this work, the 38-dimensional input vector $\mathbf{x}$ corresponds to the 38 magnetic complexity features described in~\cite{al-ghraibah2015} and summarized in Table~\ref{tab:features}, the weight parameters $\mathbf{w}$ are chosen to optimize some criterion (minimization of an $\epsilon$-insensitive error function for the case of SVR), and the output $y(\mathbf{x},\mathbf{w})$ indicates the prediction of flare size or time-to-flare.  The function $y(\mathbf{x},\mathbf{w})$ is linearly related to $\mathbf{w}$ but the transformation $\bm{\phi}(\mathbf{x})$ allows for introduction of a nonlinear relationship to the input data $\mathbf{x}$.  

\subsection{Support Vector Regression (SVR)}
Support vector machines (SVMs)~\citep{burges1998,vapnik1998,smola2004} have become popular and are considered state-of-the-art for regression problems~\citep{sapankevych2009}.  An important property of SVMs is that they can be formulated as a convex optimization problem, assuring that any local optimum is also a global optimum~\citep{bishop2006}. Support vector regression minimizes the regularized error function
\begin{equation}
 C\sum_{n=1}^{N}E_\epsilon(y_n-t_n)^2+\frac{1}{2}||\mathbf{w}||^2,
 \label{eq:svr_error}
\end{equation}
where $C$ is the inverse regularization parameter, $N$ is the total number of training points, $E_\epsilon$ is an $\epsilon$-insensitive error function, $y_n$ is the output for the $n$-th training point, $t_n$ is the target for the $n$-th training point, and $||\cdot||$ is the Euclidean norm.  The use of an $\epsilon$-insensitive error function returns a value of 0 if the absolute difference $|y_n-t_n|<\epsilon,~\epsilon>0$ and encourages sparse solutions~\citep{bishop2006}.  In this work we use the MATLAB SVM-KM toolbox~\citep{SVM-KMToolbox} and the transformation $\bm{\phi}$ implicitly defined by the radial basis (Gaussian) kernel function $k(\mathbf{x},\mathbf{x^\prime})=\bm{\phi}(\mathbf{x})^\top\bm{\phi}(\mathbf{x}^\prime)=\exp(-||\mathbf{x}-\mathbf{x}^\prime||^2/2\sigma^2)$.  The parameter $\epsilon$ is chosen based on the specific problem (size or time regression) and we optimize the values of $C$ and $\sigma$ by grid search.  

The optimal value of weight vector $\mathbf{w}$ is determined by minimizing the error function in Equation (\ref{eq:svr_error}).  This minimization is conducted using a training dataset which consists of $N$ pairs of feature vectors and labels $(\mathbf{x}_n,t_n),~n=1,\ldots,N$; the performance of the regressor is then determined on the basis of a test dataset consisting of pairs of feature vectors and labels disjoint from the training data.  The use of a disjoint test dataset is important since the optimization can overfit a solution to the training data, resulting in an inflated performance when evaluated on the training data.  Since the parameters $C$ and $\sigma$ must also be optimized, an additional dataset must be provided for this optimization; this dataset is commonly called the tuning dataset (since it is used to tune the parameters of the regressor) and is disjoint from both the training and testing datasets.

\subsection{Support Vector Regression in Time Series Analysis}
Support vector regression has been shown to outperform other nonlinear methods, including multi-layer perceptrons, for time series prediction~\citep{sapankevych2009}.  Time series prediction algorithms have been used in many practical applications such as financial market prediction (e.g., \citet{tay2001}), electric utility load forecasting~(e.g., \citet{elattar2010}), weather and environmental prediction (e.g., \citet{wang2003,mohandes2004,yu2006,wu2008}), biomedical signal processing for estimation of missing microarray data (e.g., \citet{wang2006}), and travel time prediction (e.g., \citet{wu2004}). In the aforementioned references, there is no significant feature extraction and the focus is generally on prediction using one-dimensional time series data.  Image deblurring is studied in~\citet{li2007} in which the image regions are reformulated as one-dimensional signals prior to application of SVR. To the best of our knowledge, there are no studies of SVR applied to the problem of estimating a continuous label vector from a high-dimensional image feature vector, nor of SVR applied to the prediction of solar flare characteristics. 

\section{Flare Size and Time Prediction}
\label{sec:data_experiment}
\subsection{Dataset}
\label{sec:data}
In our previous work~\citep{al-ghraibah2015}, we used a large set of line-of-sight Michelson Doppler Imager (MDI) magnetograms from 2000 through 2010. This included NOAA ARs 8809--10933, and some 594,000 total images. We describe the complexity of each AR by extracting a large set of features of postulated importance for the build up of magnetic energy and compare these to the onset of solar flares over a range of time periods following each magnetogram. We consider three general categories of features:  (1) Snapshots in space and time, encompassing total magnetic flux, magnetic gradients, and neutral lines, (2) Evolution in time, encompassing flux evolution, and (3) Structures at multiple size scales, encompassing wavelet analysis.   In total, 38 different spatial features are extracted from each magnetogram image as summarized in Table~\ref{tab:features} and as described in detail in~\citet{al-ghraibah2015}.   In this work, we predict the solar flare sizes or predict the time to flare occurrence (when in the future a flare is expected to occur) by redefining the label vector to be continuous rather than discrete. These continuous label vectors have continuous numbers related to the flare size or time of flare at each flaring region. 

\subsection{Error Measures}
Here we describe and define the error measures we use for quantifying the performance of the regression models. In the following equations, $t$ denotes the actual target value, $\hat{t}$ is the target value estimated by regression, and $N$ is the total number of data points.  For notational simplicity, we drop the index over $n=1,\ldots,N$ and interpret any summations as being over this range.

The root mean square error (RMSE) measures the square root of the mean square error:
\begin{equation}
\label{eq:rmse}
RMSE =  \sqrt{\frac{\sum{(t - \hat{t})^{2}}}{N}}.
\end{equation}
The mean absolute error (MAE) is given by:
\begin{equation}
MAE =  \frac{\sum{|t - \hat{t}|}}{N}.
\end{equation}
Both the RMSE and MAE are expressed in the same units as the original target variable $t$ and give a measure of the unsigned error.

The relative mean error (RME) is defined as:
\begin{equation}
RME =  \frac{\sum{(t - \hat{t})}}{N}.
\end{equation}
The RME is a signed error which can give a measure of the bias of the regressor, i.e., whether we consistently over- or under-estimate the target variable $t$.  Positive values of RME indicate a systematic underestimation while negative value indicate a systematic overestimation of target variables.  An RME of zero indicates no systematic bias in error.

The Pearson's correlation coefficient is given by:
\begin{equation}
R =  \frac{\sum{\left(t-\bar{t}\right)\left(\hat{t}-\bar{\hat{t}}\right)}}{\sqrt{\sum{\left(t-\bar{t}\right)^{2}} \sum{\left(\hat{t}-\bar{\hat{t}}\right)^{2}}}},
\end{equation}
where $\bar{t}$ is the mean of $t$ over the $N$ data points, and $\bar{\hat{t}}$ is the  mean of the estimates $\hat{t}$ over the $N$ data points. $R$ measures the linear correlation between the actual flare sizes and predicted (regressed) flare sizes, giving a value of $+1$ if the actual and predicted values are perfectly correlated, 0 if they are completely uncorrelated, and $-1$ if they are perfectly negatively correlated.

\subsection{Experimental Setup}
\label{sec:experiment}
\subsubsection{Size Regression}
In our size regression experiments, the goal is to predict the size of a future solar flare using SVR.  We consider two different experiments for size regression.  In Experiment 1 we apply SVR only to those regions which ultimately flare, ignoring the non-flaring regions.  This will give the most accurate results of the discriminative potential of magnetogram complexity features for determining solar flare size.  We note that in an operational scenario, an automated classification algorithm (such as that presented in~\citet{al-ghraibah2015}) could be used to automatically separate flaring from non-flaring regions prior to size regression.  In Experiment 2 we apply SVR to both flaring and non-flaring regions, as well as consider a thresholding of the regression results for flare prediction.

Solar flares are categorized as A-, B-, C-, M- or X-class according to the peak flux (in watts per square meter, W m$^{-2}$) of 100--800 pm X-rays near Earth, as measured by the \emph{geostationary operational environmental satellite} (\emph{GOES}). As reference, Table~\ref{tab:flare_sizes} shows the solar flare classes and their corresponding peak fluxes.  These \emph{GOES} classes are the standard for measuring absolute size of flares; however, due to the exponential nature of the class definitions, GOES classes are not convenient for measuring the error of a size estimate.  For example an error of $10^{-5}$ W m$^{-2}$ has a very different significance when interpreted in light of a C-class flare than an X-class flare.  As such, we seek a continuous label corresponding to flare size for which a given error can be interpreted independent of the underlying flare class.  This would imply, for example, that an error of 1 arbitrary units would indicate a prediction of a C2.0 rather than a C1.0 \emph{or} an X2.0 rather than an X1.0.

Rather than an exponentially spaced label vector, we define a linearly spaced continuous-valued label vector for flare size as follows and as summarized in Table~\ref{tab:flare_sizes}.  Flare magnitudes are offset by 0 for B-class, 9 for C-class, 18 for M-class, and 27 for X-class.  For example, a C3.3 flare would be assigned value $3.3+9=12.3$ and an M5.7 would be assigned value $5.7+18=23.7$.  We note that the two A-class flares in our dataset are assigned fractional values close to 1 and are aggregated into the B-class for data balancing (discussed below).  We define this linearly spaced continuous-valued label vector for flare size for a range of different predictive time windows. Thus, for a predictive time window of $k$ hours, if an AR flares within $k$ hours of the present time, we define the label as the linearly spaced flare size. Since the predictive time window is larger than the nominal cadence of the MDI data (96 minutes), a single flare event will be regressed independently by multiple feature vectors.  Furthermore, it is possible that more than one flare occurs in the given predictive time window.  In such a case, we define the corresponding entry in the label vector to be the largest flare size which occurs in that time window.  We repeat this experiment for $k$ in the range $[2,24]$ hours in steps of 2 hr. The error measures are computed to compare the estimated flare sizes with the real flare sizes at each time $k$ in the time window.  We further note that the \emph{GOES} background level will have an effect on the accuracy of the regression for Experiment 2 when non-flaring regions are included.  If a flare is predicted to be smaller than the concurrent \emph{GOES} background flux, the label computed from the \emph{GOES} flare catalog will indicate no flare and we will see an apparent error equal to the prediction flare size; it is possible, however, that a flare actually did occur but was missed by the \emph{GOES} instrument due to the high background.

Since the distribution of flares is unbalanced across the classes (more smaller flares than larger), we balance the data by randomly subsampling the larger classes to the size of the smallest class to build a more accurate regression model.  The class with smallest number of data samples will be the X-class flares in our dataset.  We use 100-fold cross validation for each time $k\in[2,24]$.  In other words, we randomly subsample the data 100 times and train a regressor on each of these 100 data subsets; furthermore, each of these 100 data subsets is balanced such that there is equal representation among the flare classes.  Since the data balancing inherently removes a large portion of the training data, cross validation in this scenario allows us to better utilize our entire training dataset by considering multiple randomly chosen subsets and averaging the results.  We choose 100-fold cross validation since the number of X-class flares is approximately 1\% of the total number of flares in the worst case; this means that, statistically, we will use each data sample at least once in the training process.

We choose a value of $\epsilon=0.1$ which will penalize errors greater than a tenth of a magnitude; we choose this value on the basis of the fact that the \emph{GOES} magnitudes are specified only to a significant digit of a tenth.  The SVR inverse regularization parameter $C$ controls the smoothness of the decision boundary and the kernel parameter $\sigma$ controls the smoothing of the input data; both $C$ and $\sigma$ are optimized for $k=24$ hr, 10-fold cross validation, and a tuning dataset consisting of a randomly chosen fourth of the dataset.  A time window of $k=24$ hr is used since this is the dataset with the largest number of X-class flares, and 10-fold cross validation is used rather than 100-fold cross validation for computational reasons.  No noticeable difference in average performance was noted with 10-fold versus 100-fold cross validation simulations.  

\subsubsection{Time Regression}
In our time regression experiments, the goal is to predict how many hours in the future a flare will occur using SVR.  In Experiment 3 we apply SVR only to those regions which ultimately flare.  For this experiment we define a label vector containing the number of hours to the next flare in an AR, as measured from the  current time.  Since there is no distinguishing between flare sizes in this experiment, there is no need for data balancing.  We choose a value of $\epsilon=1.6$ which corresponds to the 96 minute cadence of the MDI data expressed in hours.  The SVR inverse regularization parameter $C$ controls the smoothness of the decision boundary and the kernel parameter $\sigma$ controls the smoothing of the input data; both $C$ and $\sigma$ are optimized using 10-fold cross validation and a \textbf{tuning dataset consisting of a} randomly chosen fourth of the data.

\section{Results and Discussion}
\label{sec:results}
\subsection{Size Regression}
\label{sec:size_results}
In this section, we present results for two different definitions of the basic flare size label vector as defined in Table~\ref{tab:flare_sizes}.  In Experiment 1 we use the definitions for the flare label vector directly as summarized in Table~\ref{tab:flare_sizes}.  Experiment 1 is designed to study the discriminatory potential of the 38 magnetogram complexity measures for determining flare size; we note that in an operational scenario, the AR could first be classified as flaring by a separate classifier (e.g.,~\citet{al-ghraibah2015}).  In Experiment 2 we additionally include non-flaring regions and assign those regions a flare size of 0.  Experiment 2 is designed to combine flare prediction (i.e., will a flare of any size occur) with flare size prediction.

\subsubsection{Experiment 1: Size Regression, No Non-flaring Regions}
\label{sec:experiment1}
Parameters $C$ and $\sigma$ are optimized via grid search using a tuning dataset consisting of a randomly chosen fourth of the ARs.  For each of the 10-fold cross validation runs, this tuning dataset is randomly subsampled to balance the classes by choosing $N_{\text{min}}/2$ data points from each class for tuning training, where $N_{\text{min}}$ is the size of the smallest class (X-class flares in all cases).  All remaining unbalanced data unseen in the tuning training is used as tuning test data.  In other words, we vary parameters $C\in\{10,50,100,500,1000\}$ and $\sigma\in\{0.1,0.5,1,5,10,50,100,500,1000\}$ and train 10 regressors for each combination of $C$ and $\sigma$ using a balanced subset of the tuning dataset while testing performance using the tuning data which is disjoint from that used to train the regressor.  RMSE results are computed over the tuning test data and averaged over the 10-fold cross validations; we find a minimum RMSE of approximately 4.9 for $C=9$ and $\sigma=9$ which we use for subsequent size regression simulations in Experiment 1.  

\begin{figure}
 \epsscale{.40}
 \centering
 \subfloat[RMSE.]{\plotone{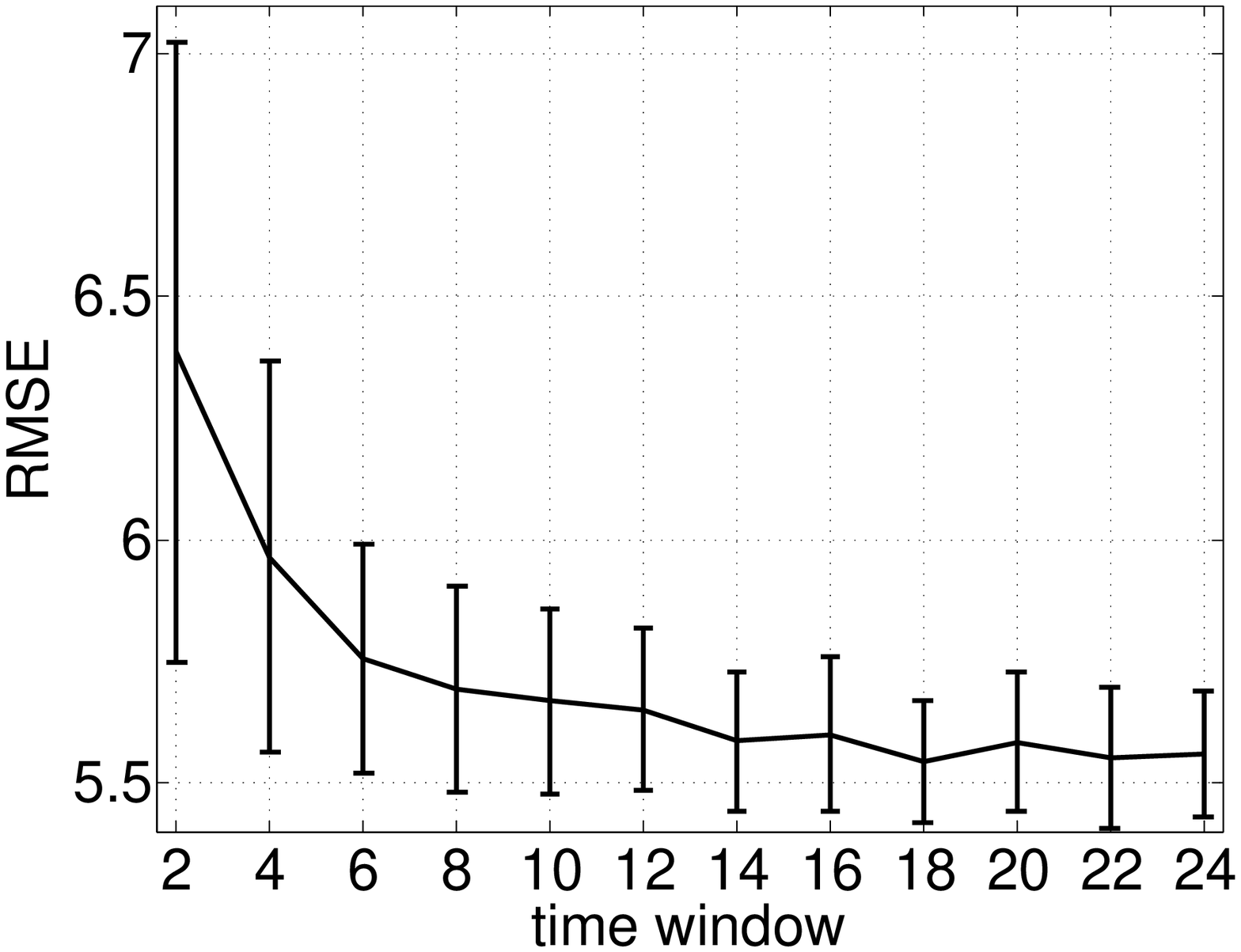}}~~
 \subfloat[MAE.]{\plotone{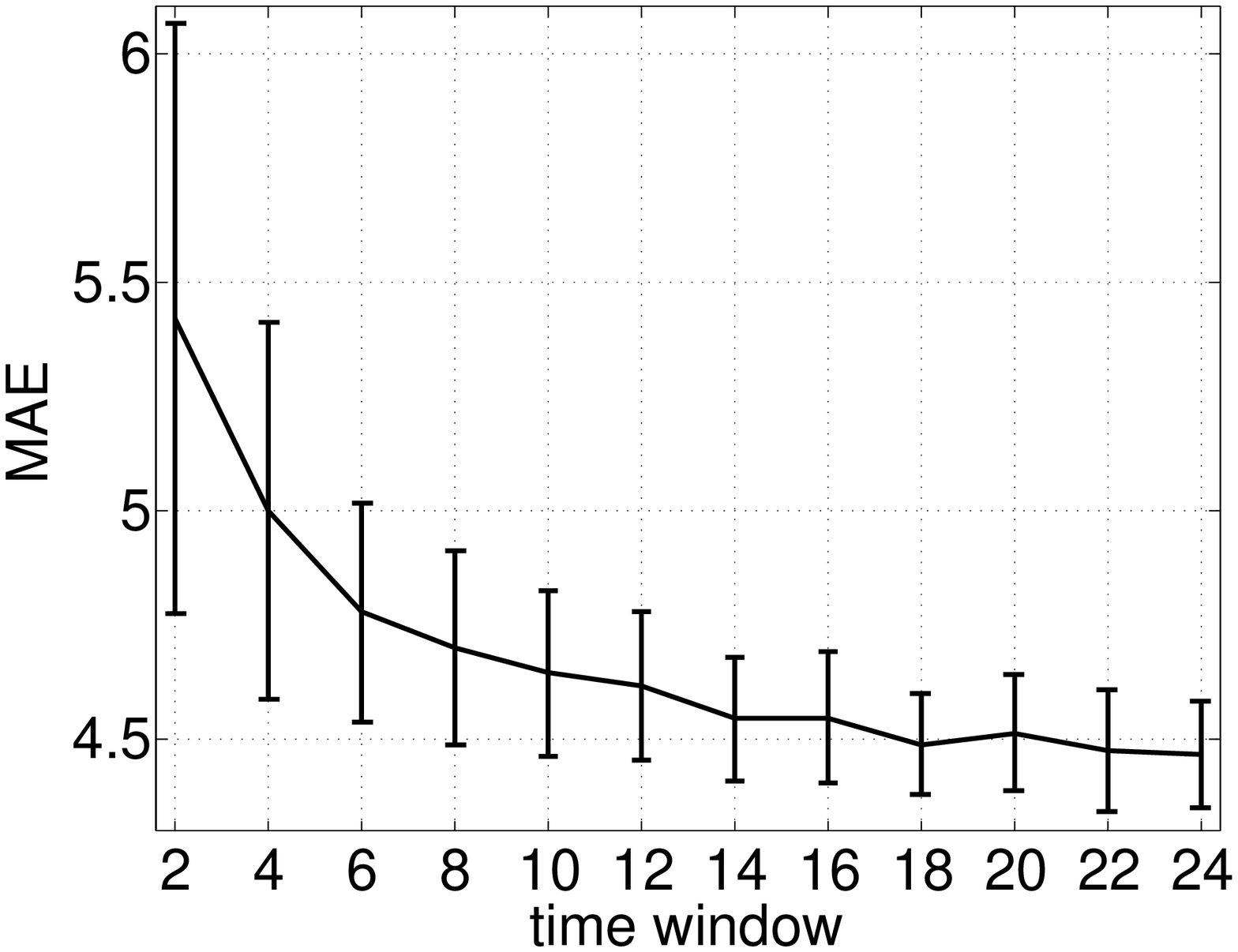}}\\
 \subfloat[RME.]{\plotone{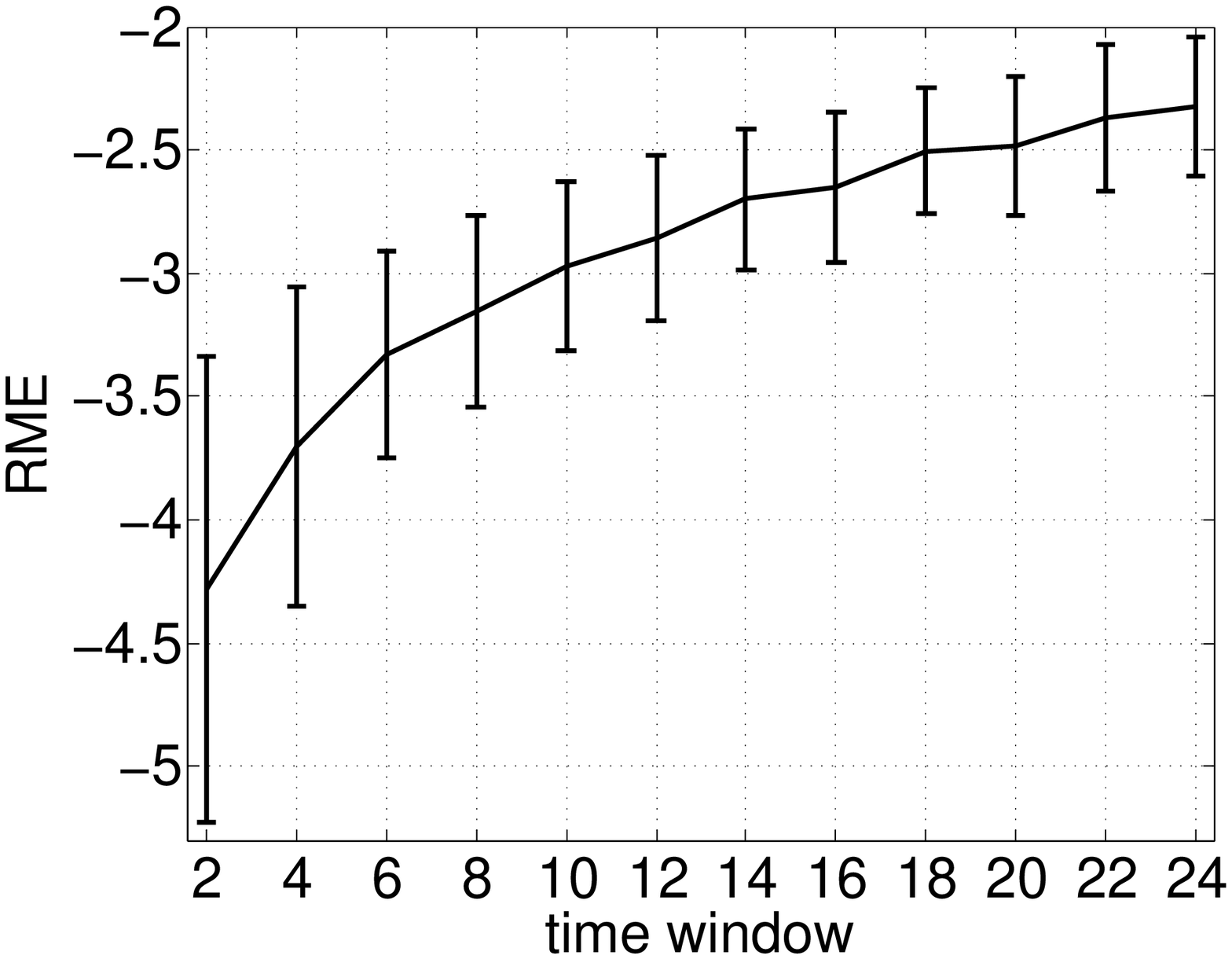}}~~
 \subfloat[Correlation coefficient.]{\plotone{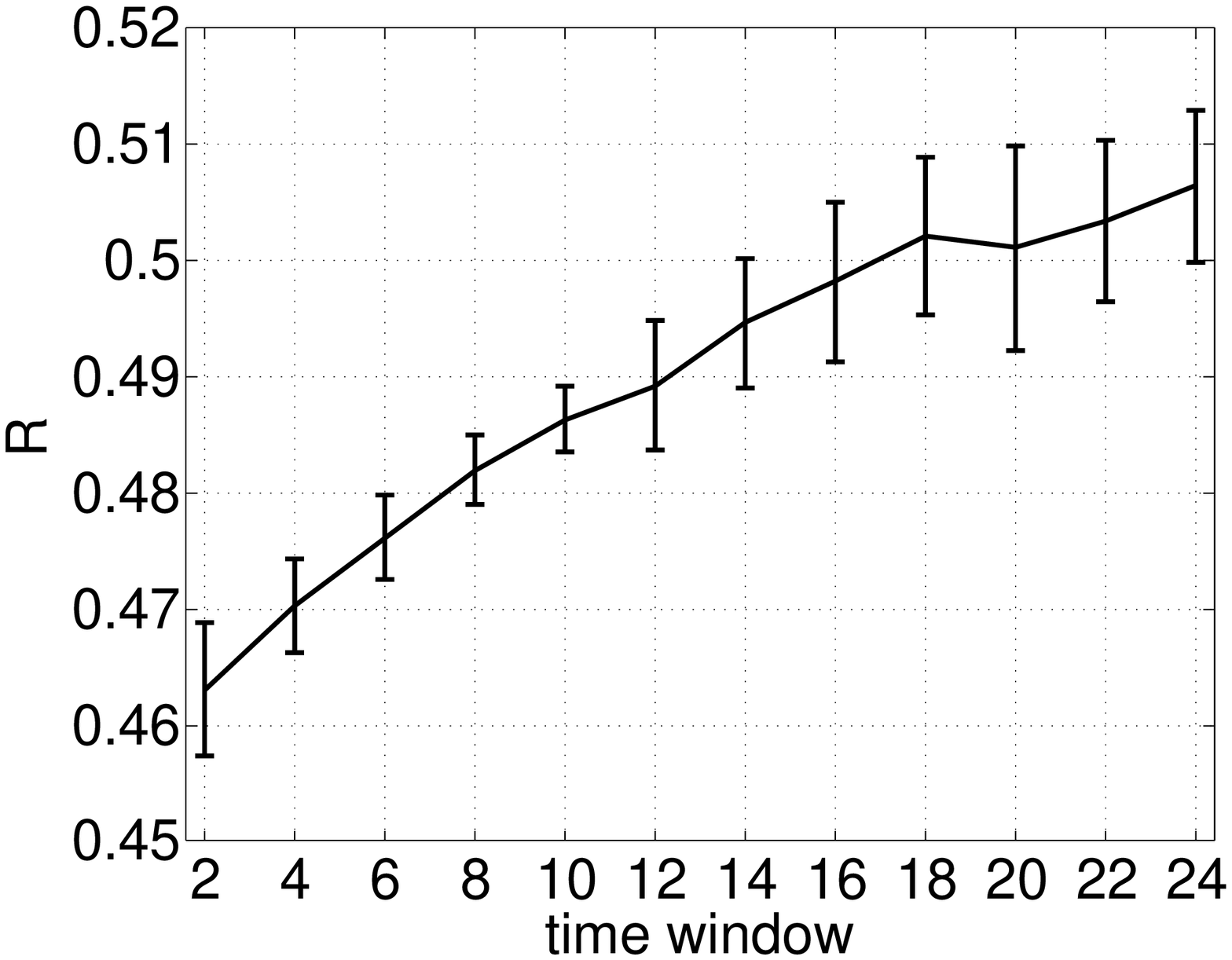}}
 \caption{Size regression error measures for Experiment 1 (size regression, no non-flaring regions). Error bars indicate the standard deviation across the 100 cross validation runs.}
 \label{fig:size_errors}
\end{figure}

We now use the remaining $3/4$ of the ARs unseen in tuning to cross validate performance of the size regression. Similar to the $C$ and $\sigma$ optimization, training data are balanced and all data unused in the balanced training are used as test data for each of the 100 folds. We plot the RMSE, MAE, RME, and correlation coefficient $R$ for size regression in Fig.~\ref{fig:size_errors}; these errors are averaged over the test data (unseen in training and tuning) for the 100-fold cross validation runs.  We see an average RMSE (Fig.~\ref{fig:size_errors}(a)) ranging between 5.6 and 6.4, with smaller errors for larger predictive time windows. With our linearly spaced flare size label vector, recall that an error of 1 indicates an error of one magnitude increment (e.g., M1.0 versus M2.0) and an error of 10 indicates an error of one class increment (e.g., M1.0 versus X1.0).  We thus find an RMSE of slightly more than half a class.  Similarly, we find an average MAE (Fig.~\ref{fig:size_errors}(b)) between 4.5 and 5.5, again with smaller errors for larger predictive time windows and again indicating an error of approximately half a class.  Looking at the RME in Fig.~\ref{fig:size_errors}(c),  we see that our estimates of flare size are biased toward overestimating by 2.3--4.3 mag increments with larger bias for smaller predictive time windows.  The correlation coefficient in Fig.~\ref{fig:size_errors}(d) ranges from 0.46 to 0.51, with larger correlation for longer predictive time windows.  

In Fig.~\ref{fig:size_errors} we have included the standard deviation across the 100-fold cross validation runs as error bars.  We note that the smaller time windows have a larger standard deviation; this is a direct result of the smaller sample sizes for the shorter time windows.  The error bars for all but the 2-hr time window are within the average values for larger time windows in RMSE and MAE.  Thus the trend of decreasing error with increasing time window may be more of an artifact of small sample statistics than underlying physics.  Even so, the consistency in error across predictive time windows may indicate a presence of magnetic complexity features correlated to eventual flare size which persist long before flaring.  It also appears that these more persistent magnetic complexity features may allow for better prediction of flare size since the bias shown in RME in Fig.~\ref{fig:size_errors}(c) is smaller with increased window length and R in Fig.~\ref{fig:size_errors}(d) is larger with increased window length.

\begin{figure}
 \epsscale{.30}
 \centering
 \subfloat[All flares, $\mu_e=-2.7$, $\sigma_e=4.9$.]{\plotone{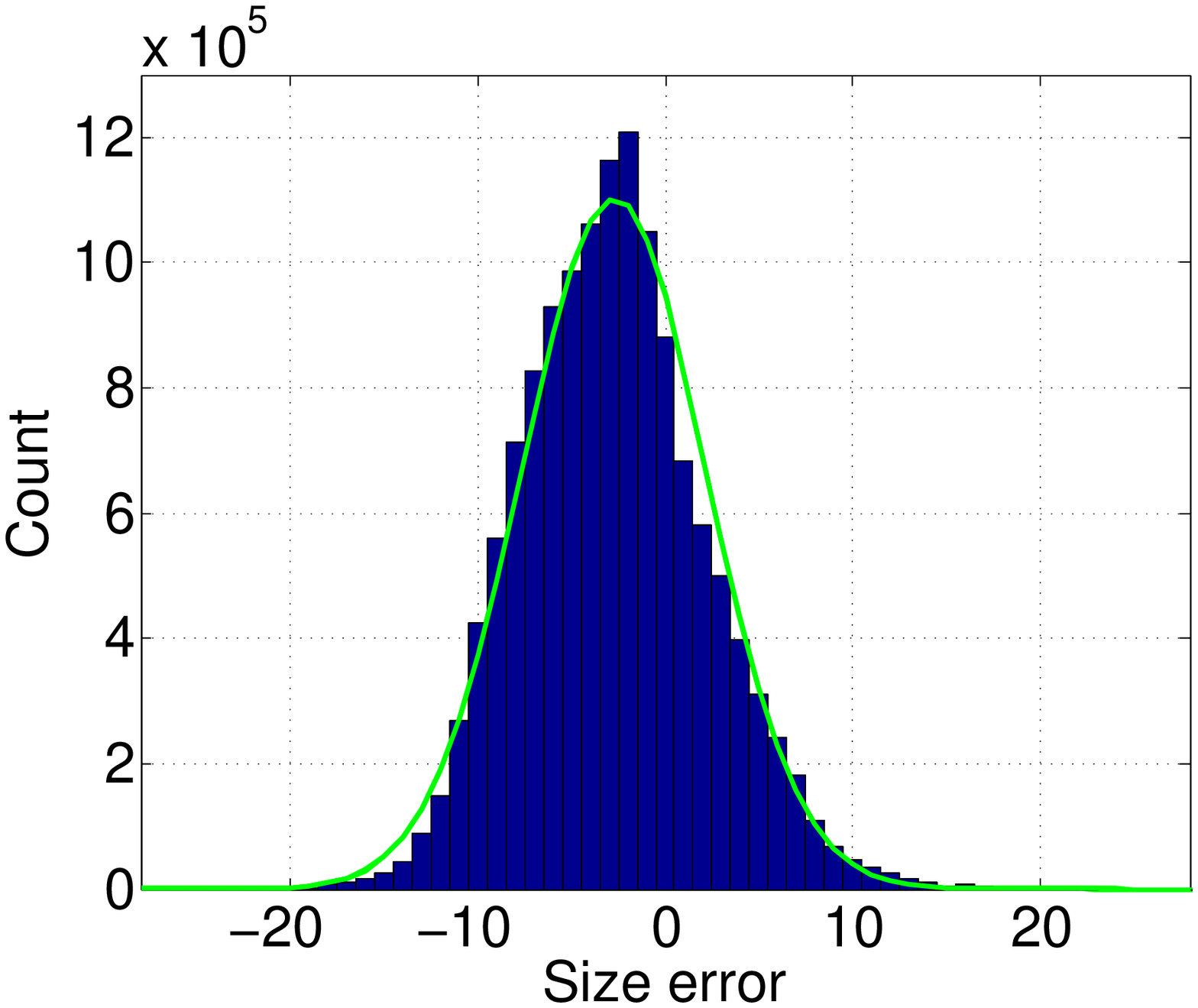}}\\
 \subfloat[A and B flares, $\mu_e=-7.3$, $\sigma_e=3.1$.]{\plotone{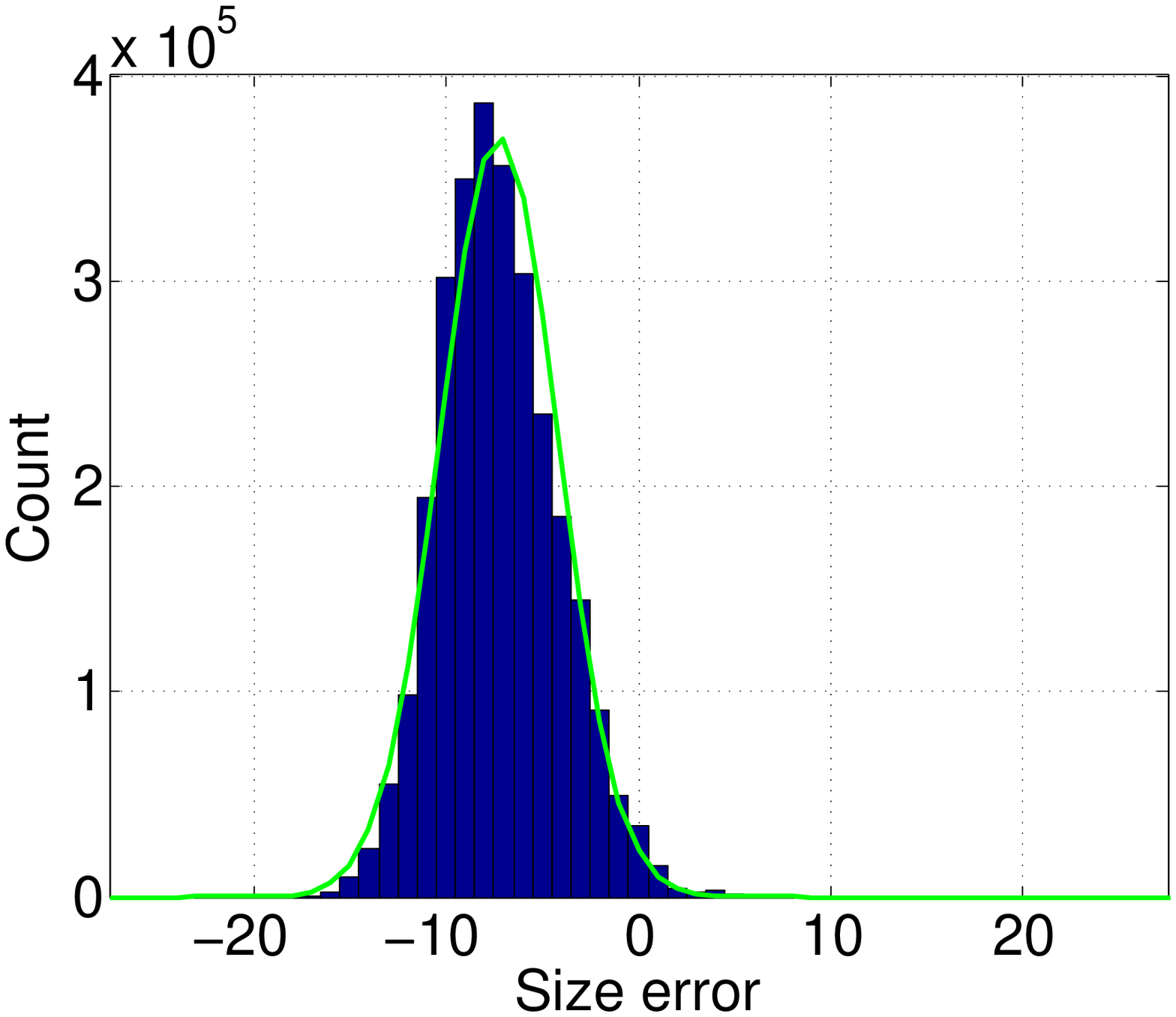}}~~
 \subfloat[C flares, $\mu_e=-2.7$, $\sigma_e=3.8$.]{\plotone{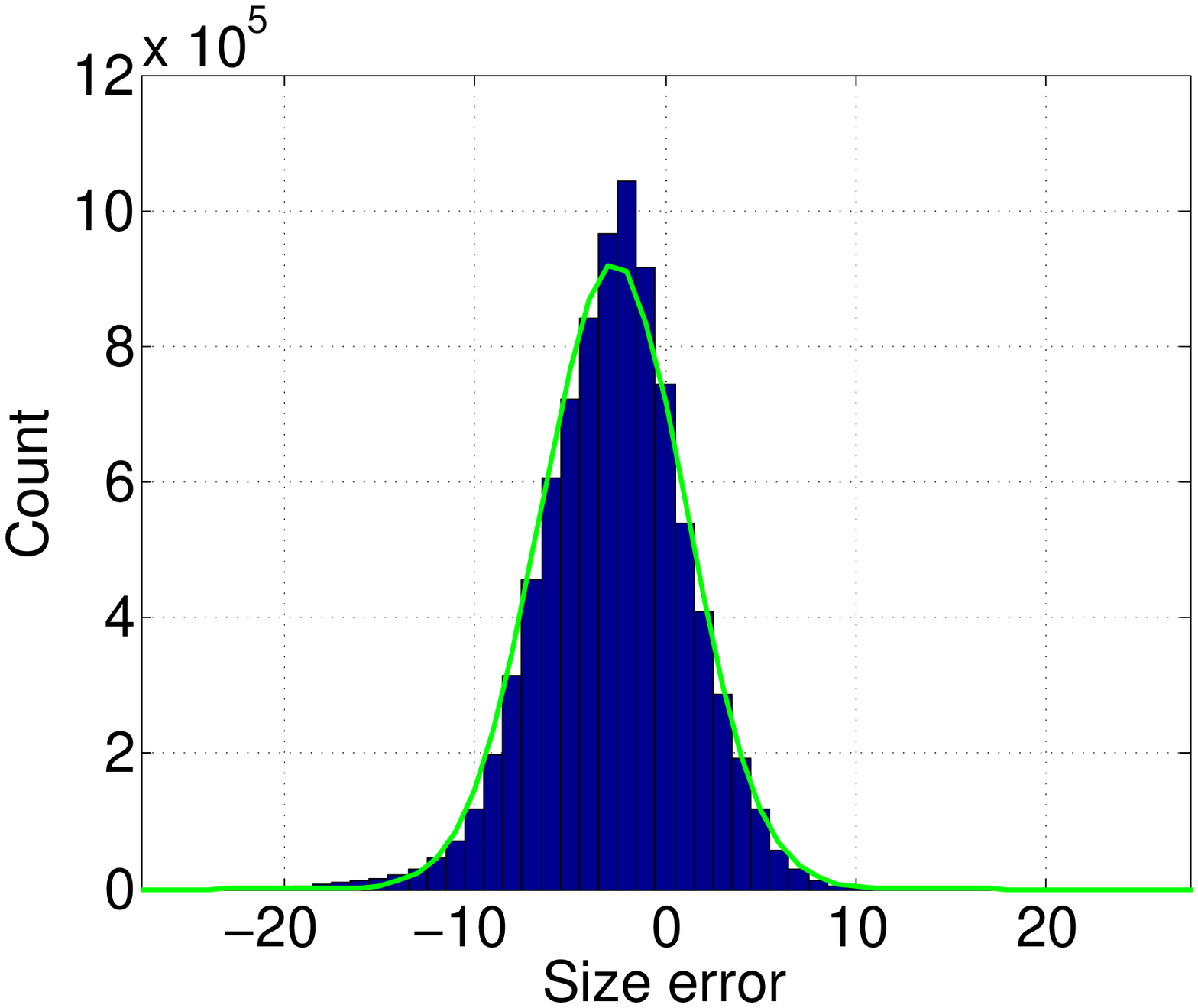}}\\
 \subfloat[M flares, $\mu_e=3.0$, $\sigma_e=4.5$.]{\plotone{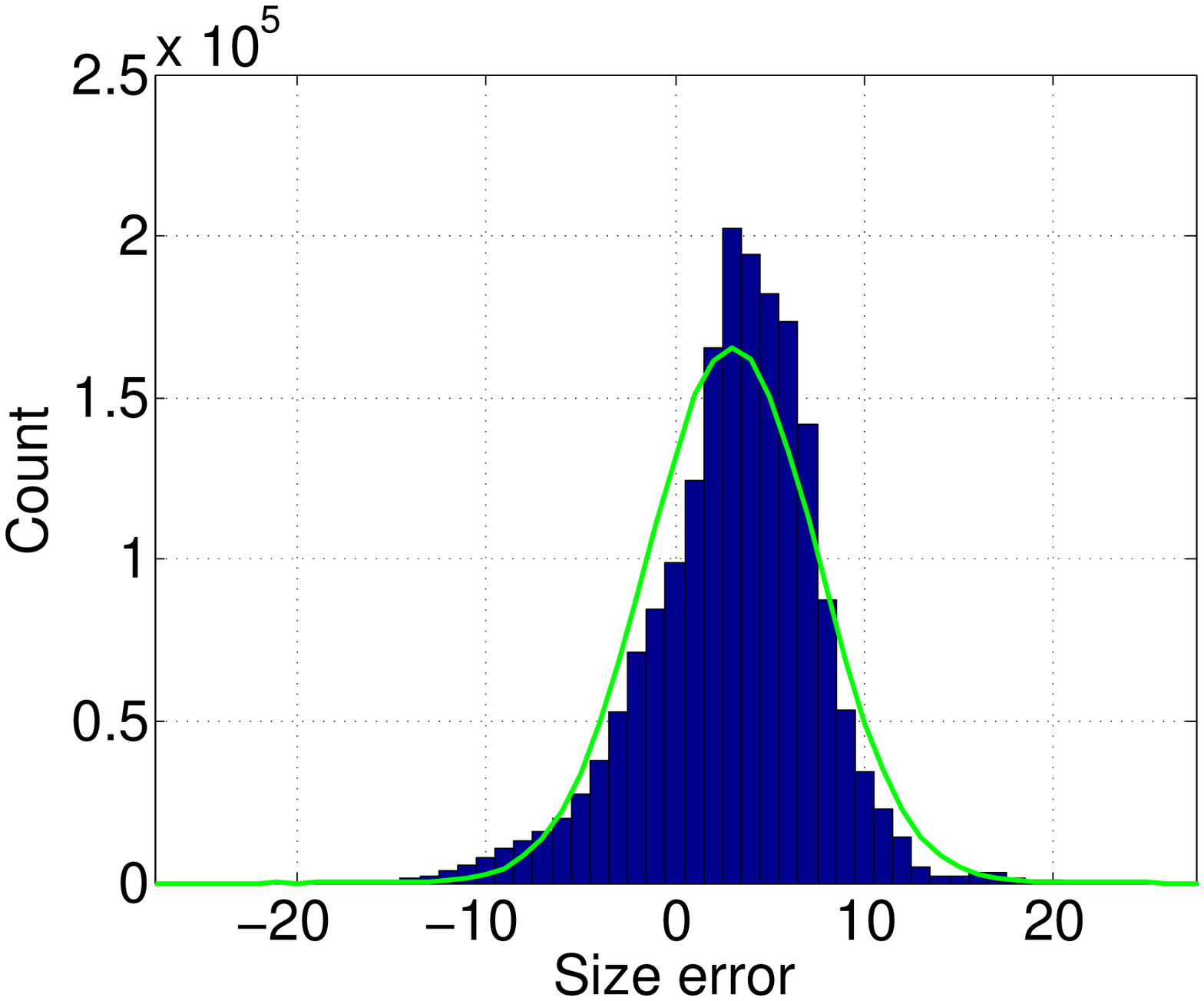}}~~
 \subfloat[X flares, $\mu_e=8.7$, $\sigma_e=4.5$.]{\plotone{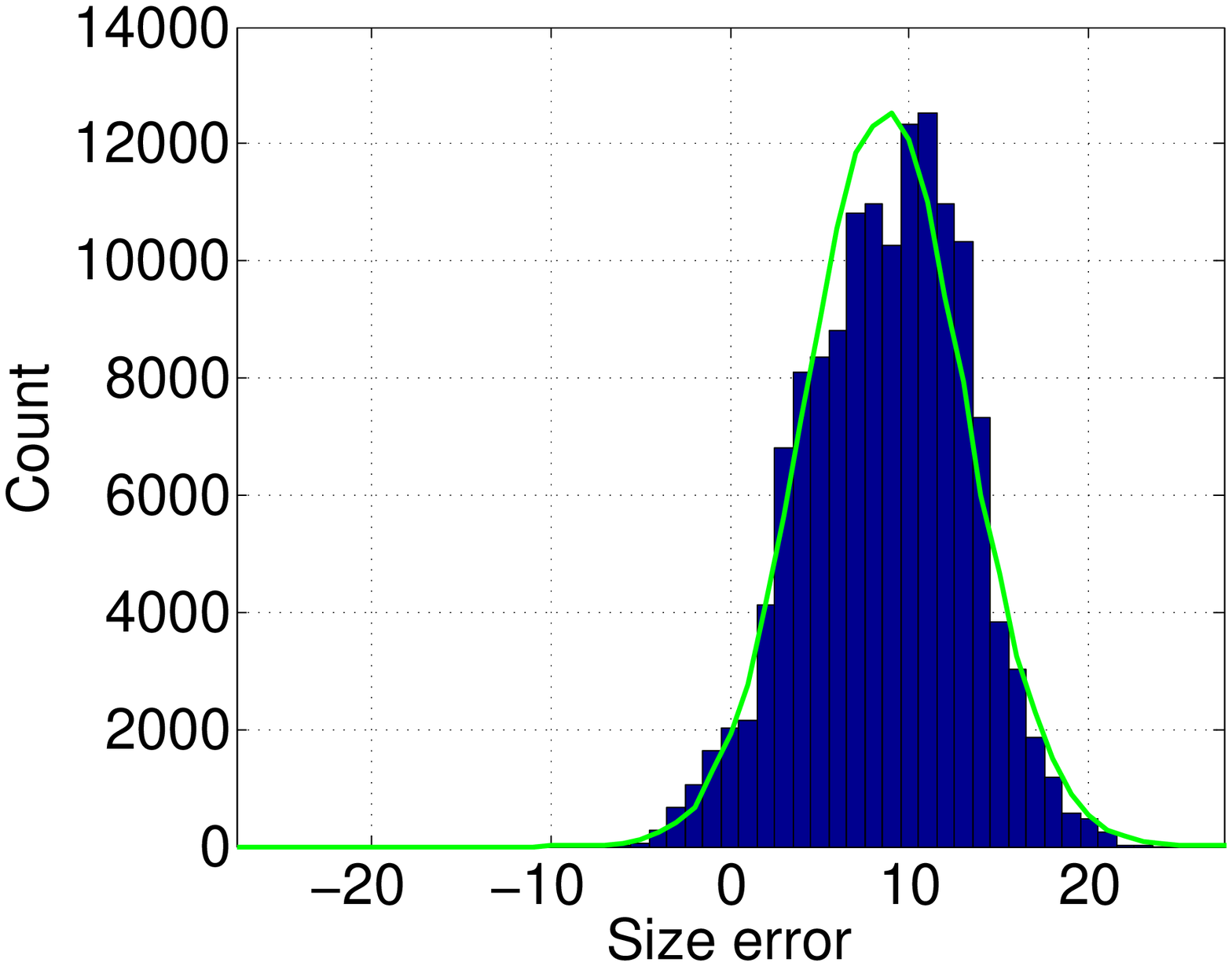}}
 \caption{Histogram of size regression error $t-\hat{t}$ for Experiment 1 (size regression, no non-flaring regions).  The sample mean $\mu_e$ and sample standard deviation $\sigma_e$ are used to plot a representative normal distribution (the green lines).}
 \label{fig:size_error_hist}
\end{figure}

In Fig.~\ref{fig:size_error_hist}(a) we plot a histogram of the error $t-\hat{t}$, aggregated over the 12 predictive time windows and the 100-fold cross validation runs.  We see that the distribution is approximately normal with a mean skewed to -2.7 (as would be expected from the RME plot in Fig.~\ref{fig:size_errors}(c)) and a standard deviation of 4.9.  As reference, the green line in Fig.~\ref{fig:size_error_hist} is a sample normal distribution with mean -2.7, standard deviation 4.9, and the same total counts as the error histogram. We additionally plot the histograms of the error $t-\hat{t}$ for each flare class in Fig.~\ref{fig:size_error_hist}(b)--(e), again aggregated over the 12 predictive time windows and 100-fold cross validation runs.  We see that the mean error is different for the different flare classes.  We find an average error of approximately 3/4 a \emph{GOES} class for the AB- and X-class flares, albeit overestimating the AB flares and underestimating the X flares.  We find a similar error of approximately 1/3 a \emph{GOES} class for the C- and M-class flares, again overestimating the smaller (C-class) flares and underestimating the larger flares.  

\subsubsection{Experiment 2: Size Regression, With Non-flaring Regions}
\label{sec:experiment2}
Parameters $C$ and $\sigma$ are optimized via grid search using the same tuning dataset and procedure as for Experiment 1, except now including non-flaring regions in the dataset and subsampling procedures.  We find a minimum RMSE of approximately 7.2 for $C=50$ and $\sigma=2$ which we use for subsequent size regression simulations in Experiment 2.  We note that the RMSE resulting from optimization of $C$ and $\sigma$ already indicates a higher average error in Experiment 2 as compared to Experiment 1.

\begin{figure}
 \epsscale{.40}
 \centering
 \subfloat[RMSE.]{\plotone{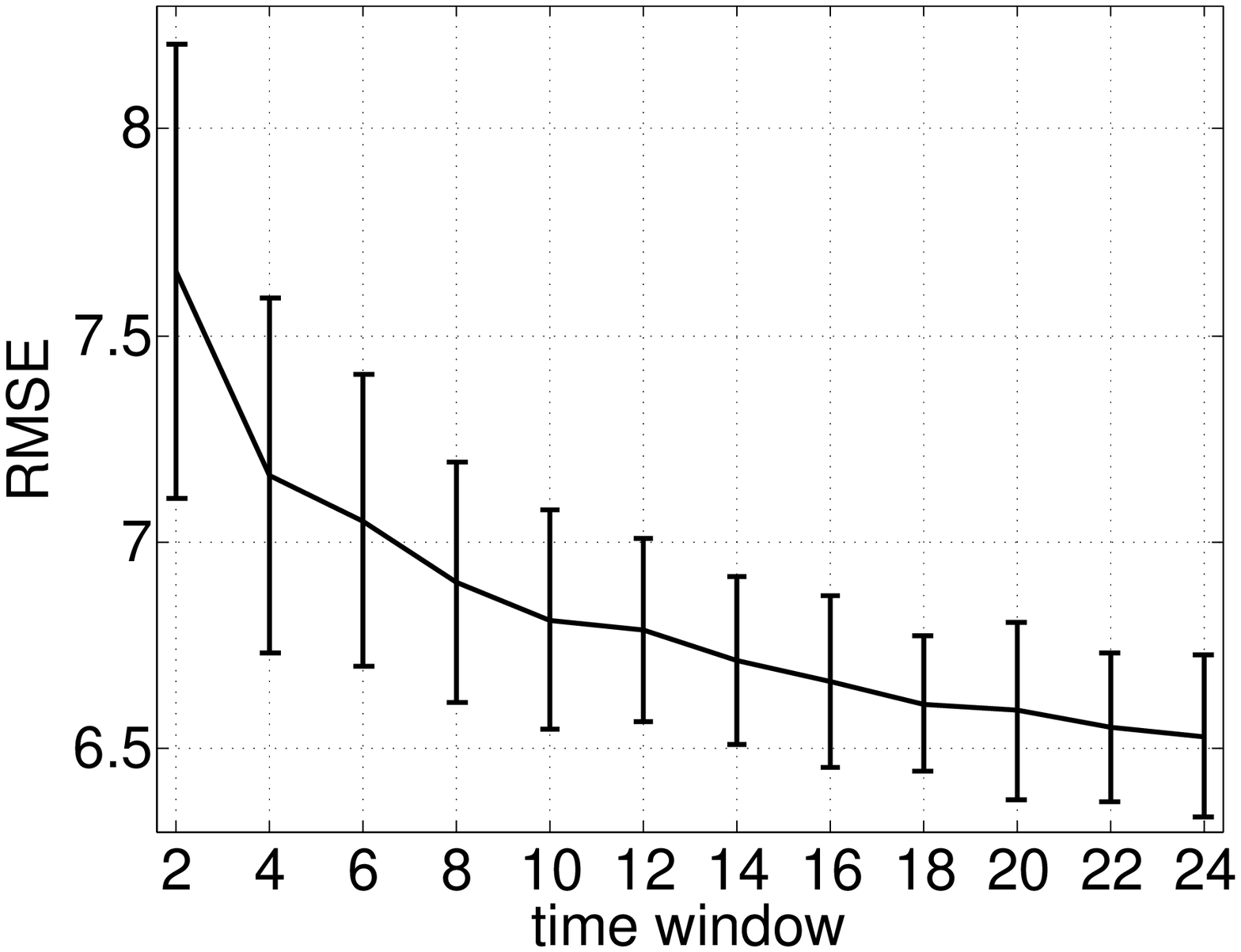}}~~
 \subfloat[MAE.]{\plotone{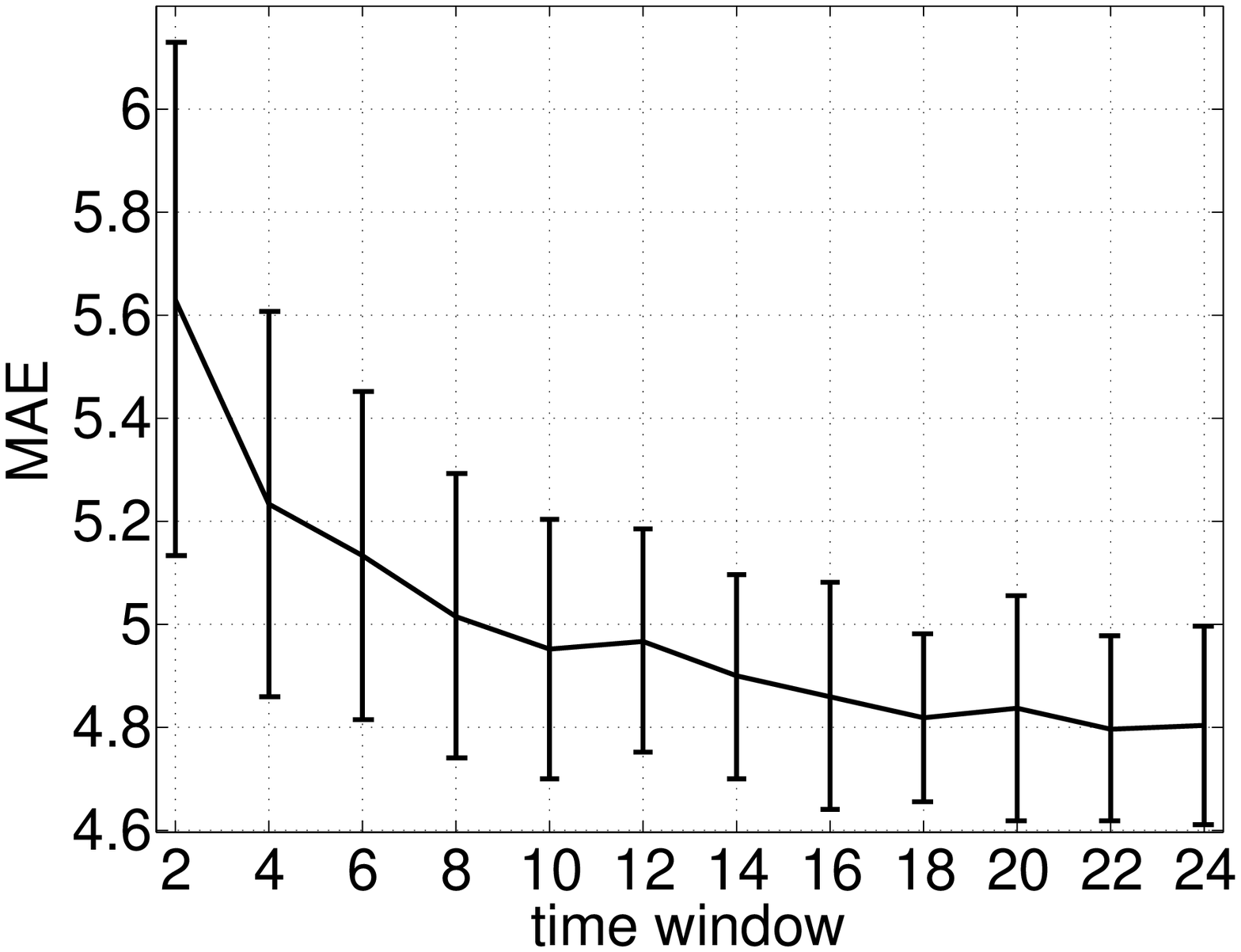}}\\
 \subfloat[RME.]{\plotone{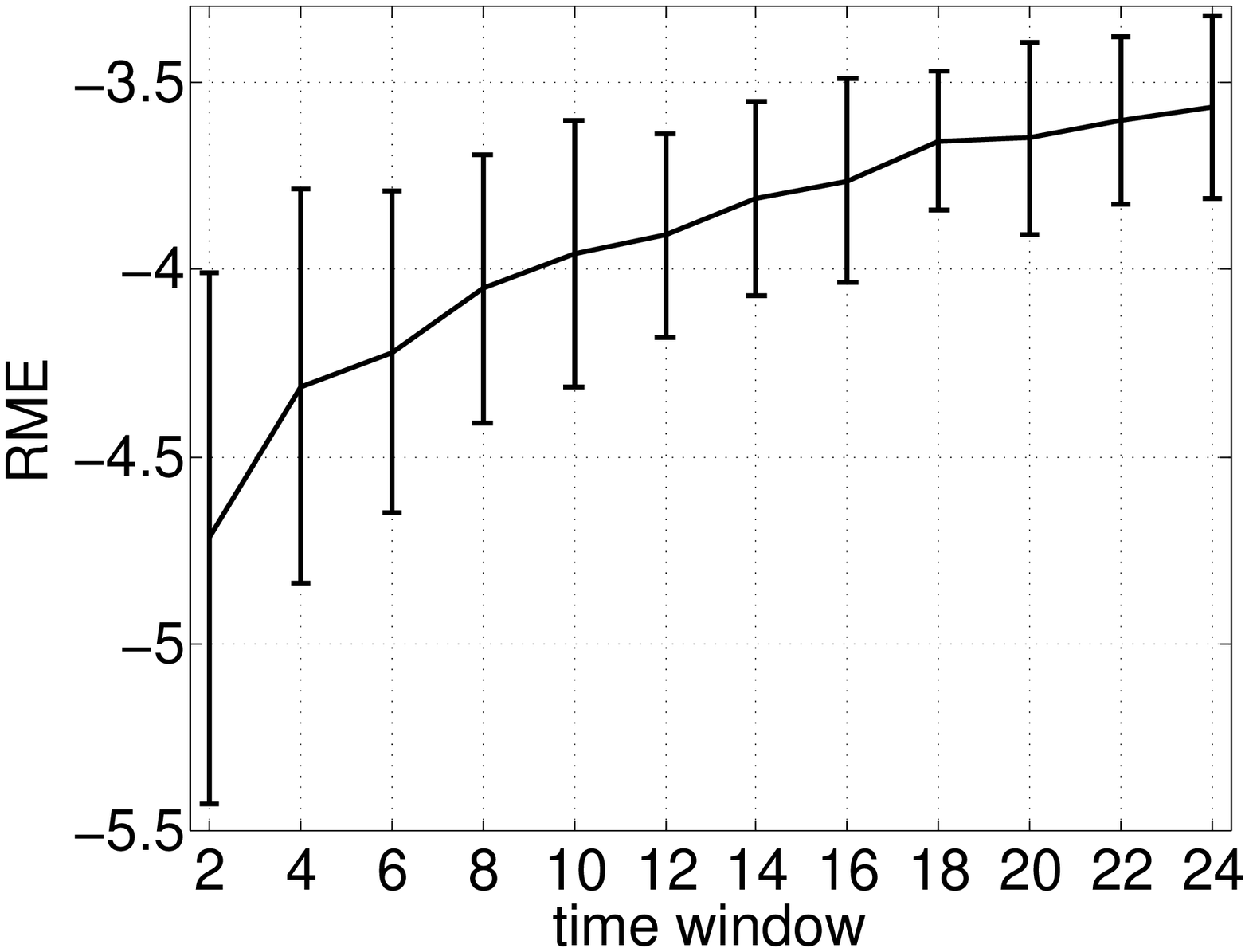}}~~
 \subfloat[Correlation coefficient.]{\plotone{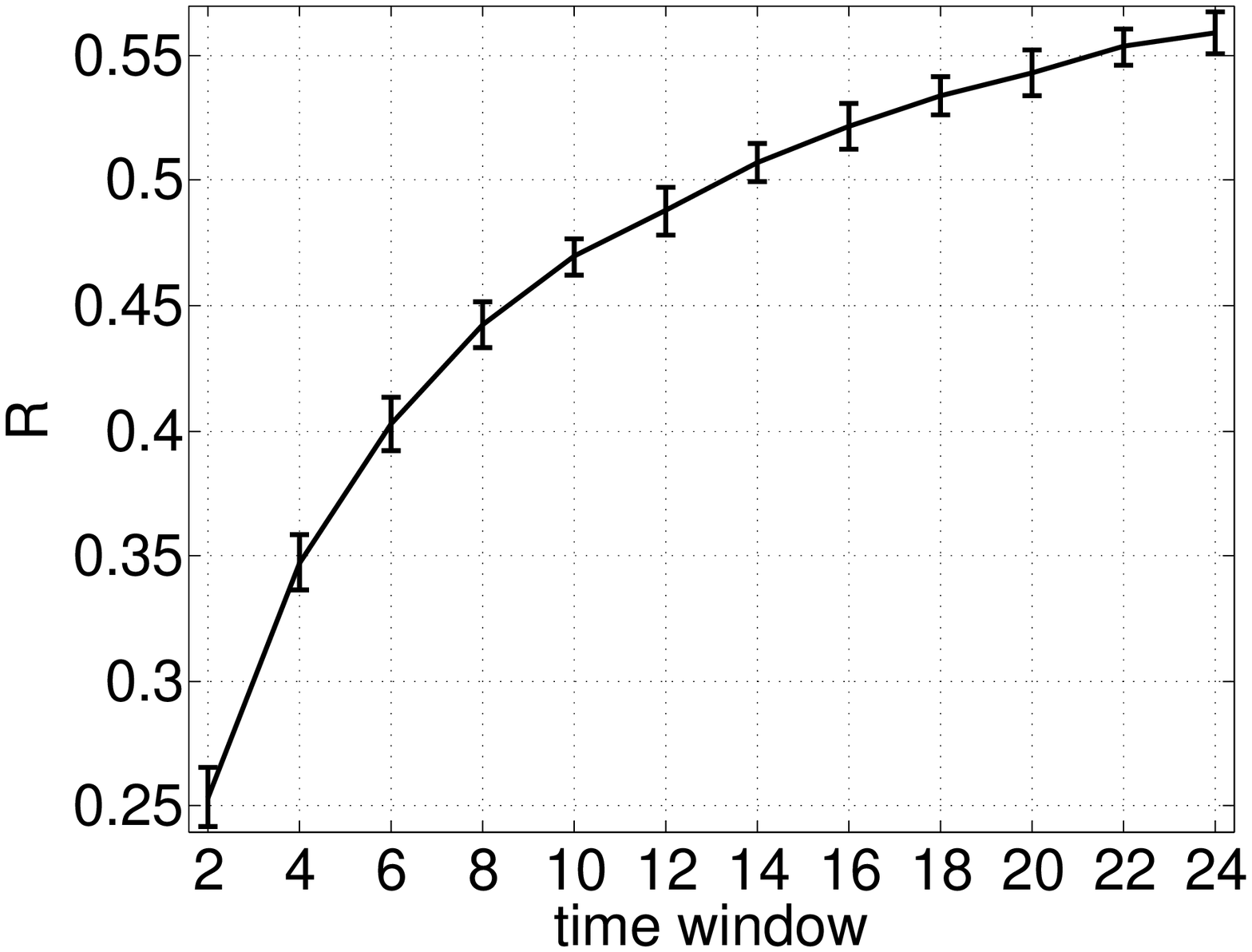}}
 \caption{Size regression error measures for Experiment 2 (size regression, with non-flaring regions).}
 \label{fig:size_errors2}
\end{figure}

We now use the remaining $3/4$ of the ARs unseen in tuning to cross validate performance of the size regression, using the same procedure as in Experiment 1. We plot the RMSE, MAE, RME, and correlation coefficient $R$ for size regression in Fig.~\ref{fig:size_errors2}.  We see an average RMSE (Fig.~\ref{fig:size_errors2}(a)) ranging between 6.5 and 7.7, with smaller errors for larger predictive time windows, corresponding to an RMSE of approximately three quarters of a class.  Similarly, we find an average MAE (Fig.~\ref{fig:size_errors2}(b)) between 4.8 and 5.6, again with smaller errors for larger predictive time windows and indicating an error of approximately half a class.  Looking at the RME in Fig.~\ref{fig:size_errors2}(c),  we see that our estimates of flare size are biased toward overestimating by 3.5--4.7 mag increments with larger bias for smaller predictive time windows.  The correlation coefficient in Fig.~\ref{fig:size_errors2}(d) ranges from 0.25 to 0.55, with larger correlation for longer predictive time windows.  In Figure~\ref{fig:size_errors2} we have included the standard deviation across the 100-fold cross validation runs as error bars.  We note that again the smaller time windows have a larger standard deviation due to small sample statistics.  

\begin{figure}
 \epsscale{.30}
 \centering
 \subfloat[All flares, $\mu_e=-3.9$, $\sigma_e=5.6$.]{\plotone{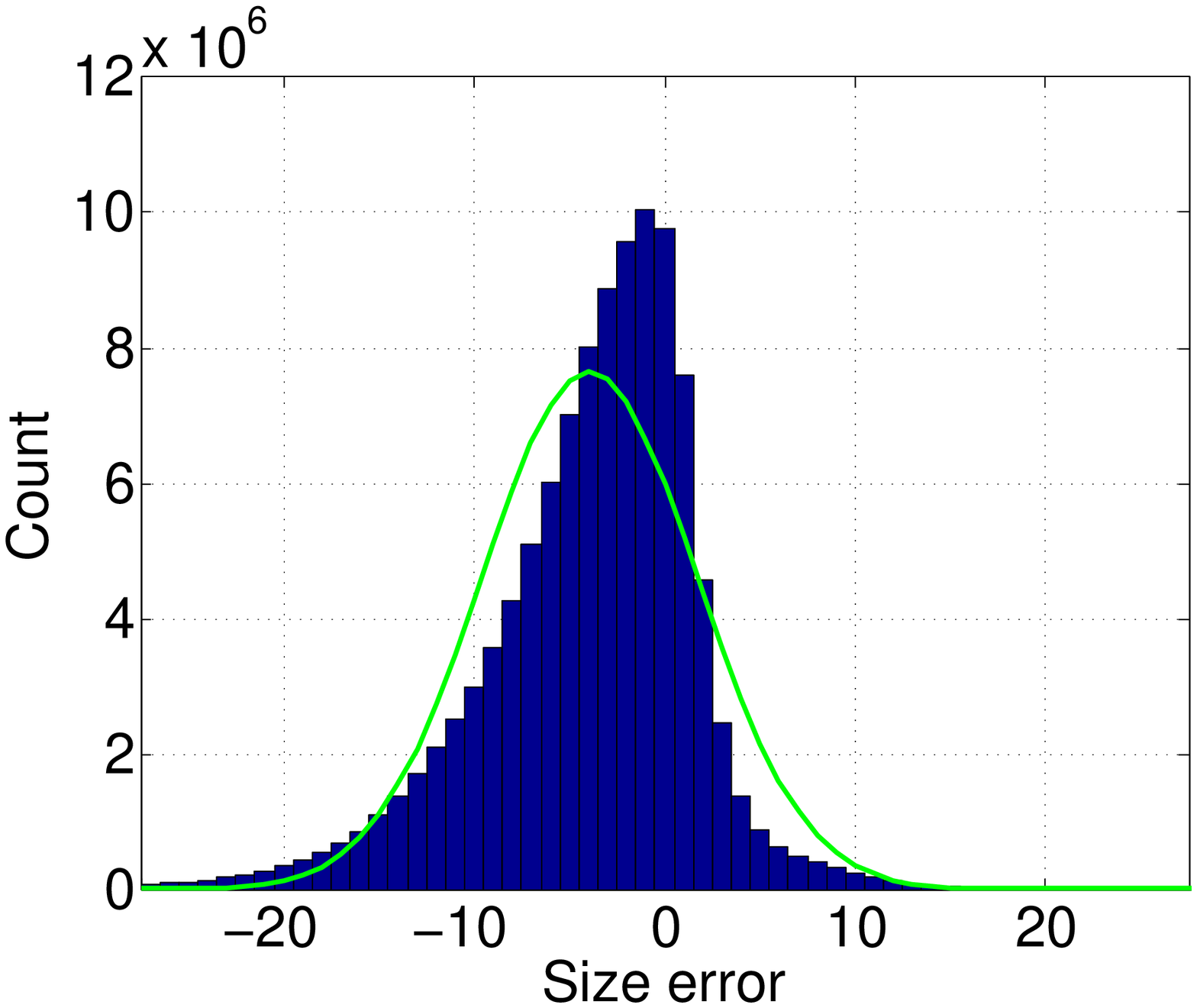}}~~
 \subfloat[Nonflaring regions, $\mu_e=-4.5$, $\sigma_e=5.3$.]{\plotone{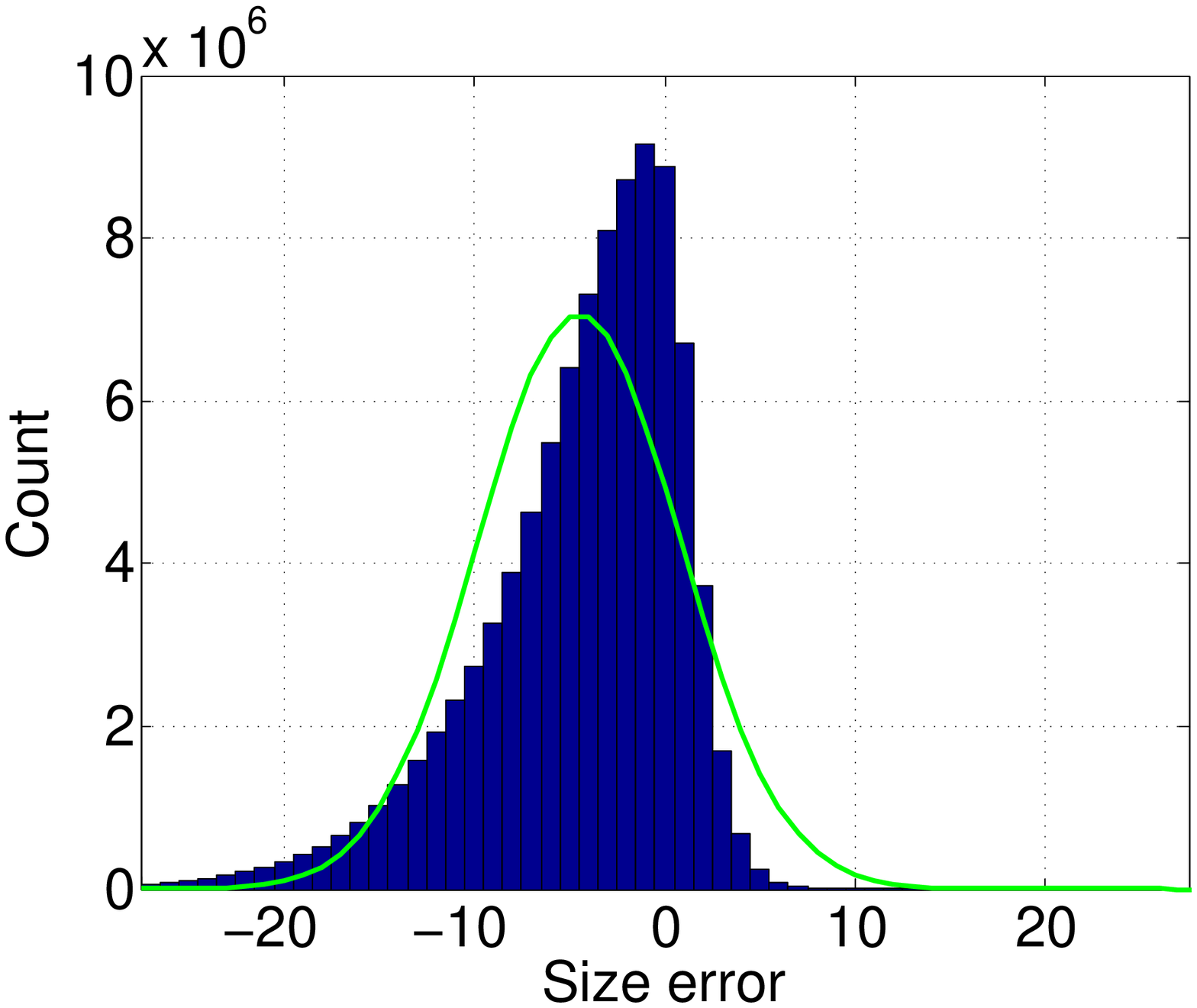}}\\
 \subfloat[A and B flares, $\mu_e=-2.4$, $\sigma_e=4.8$.]{\plotone{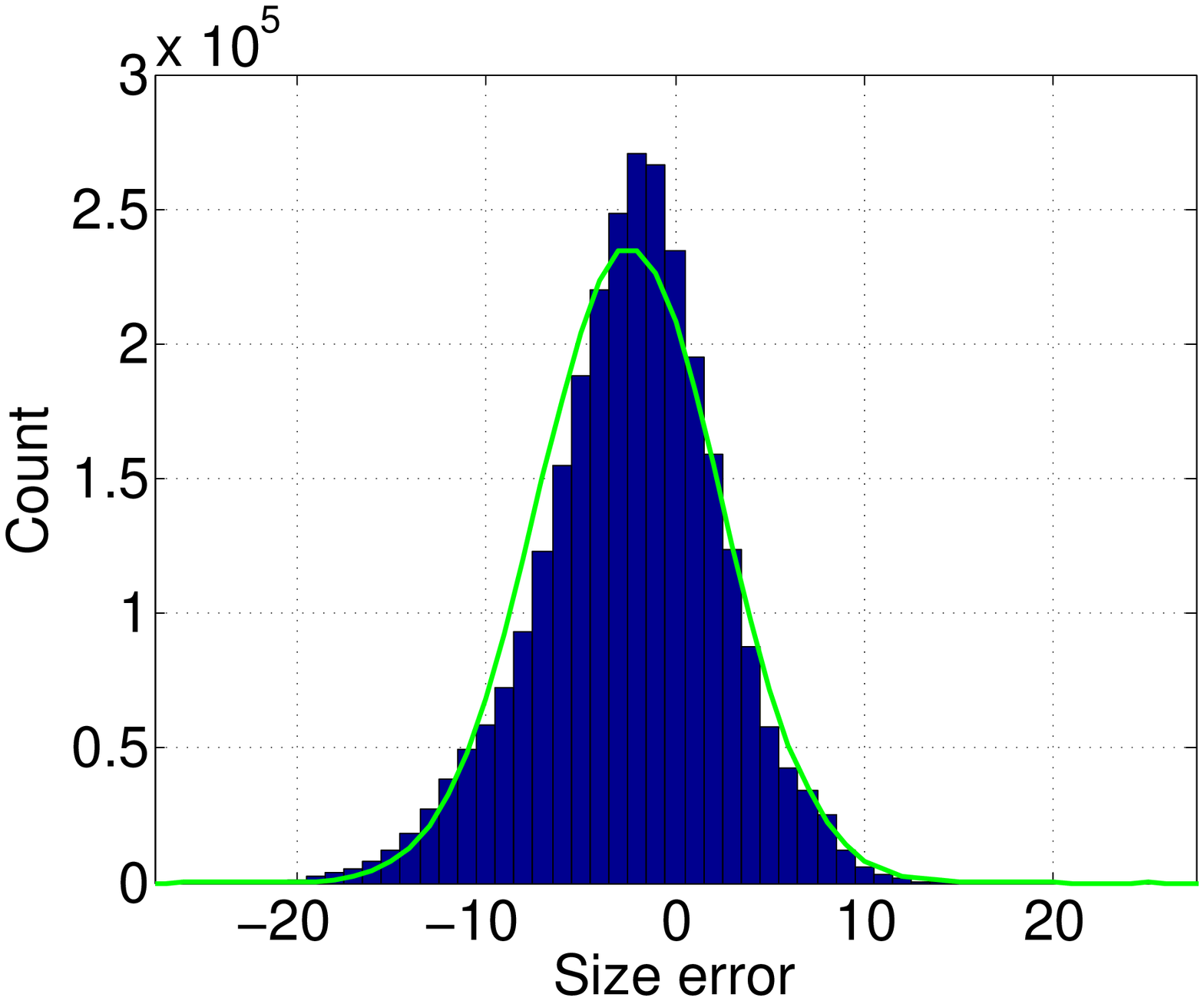}}~~
 \subfloat[C flares, $\mu_e=-0.3$, $\sigma_e=6.2$.]{\plotone{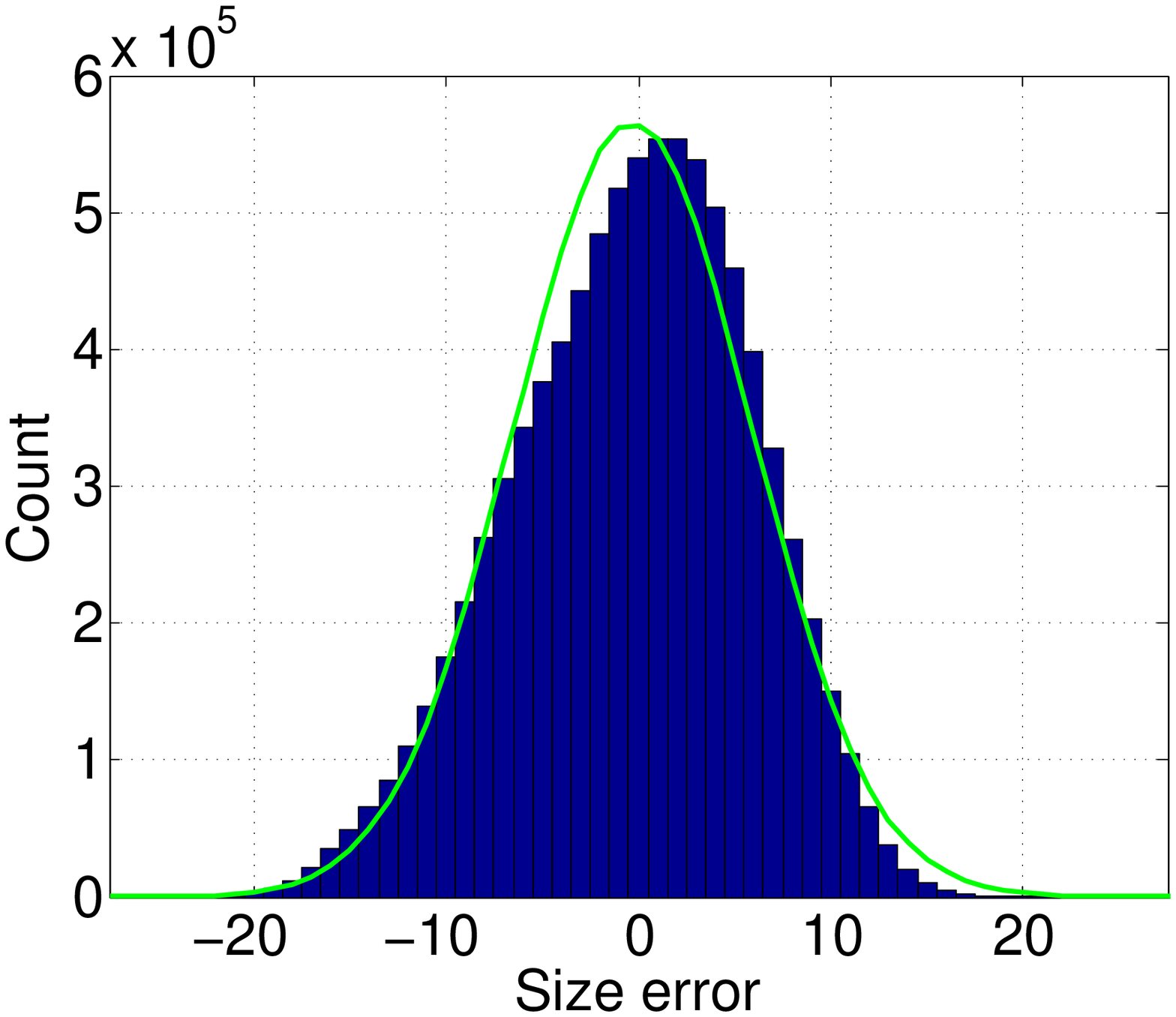}}\\
 \subfloat[M flares, $\mu_e=3.5$, $\sigma_e=6.2$.]{\plotone{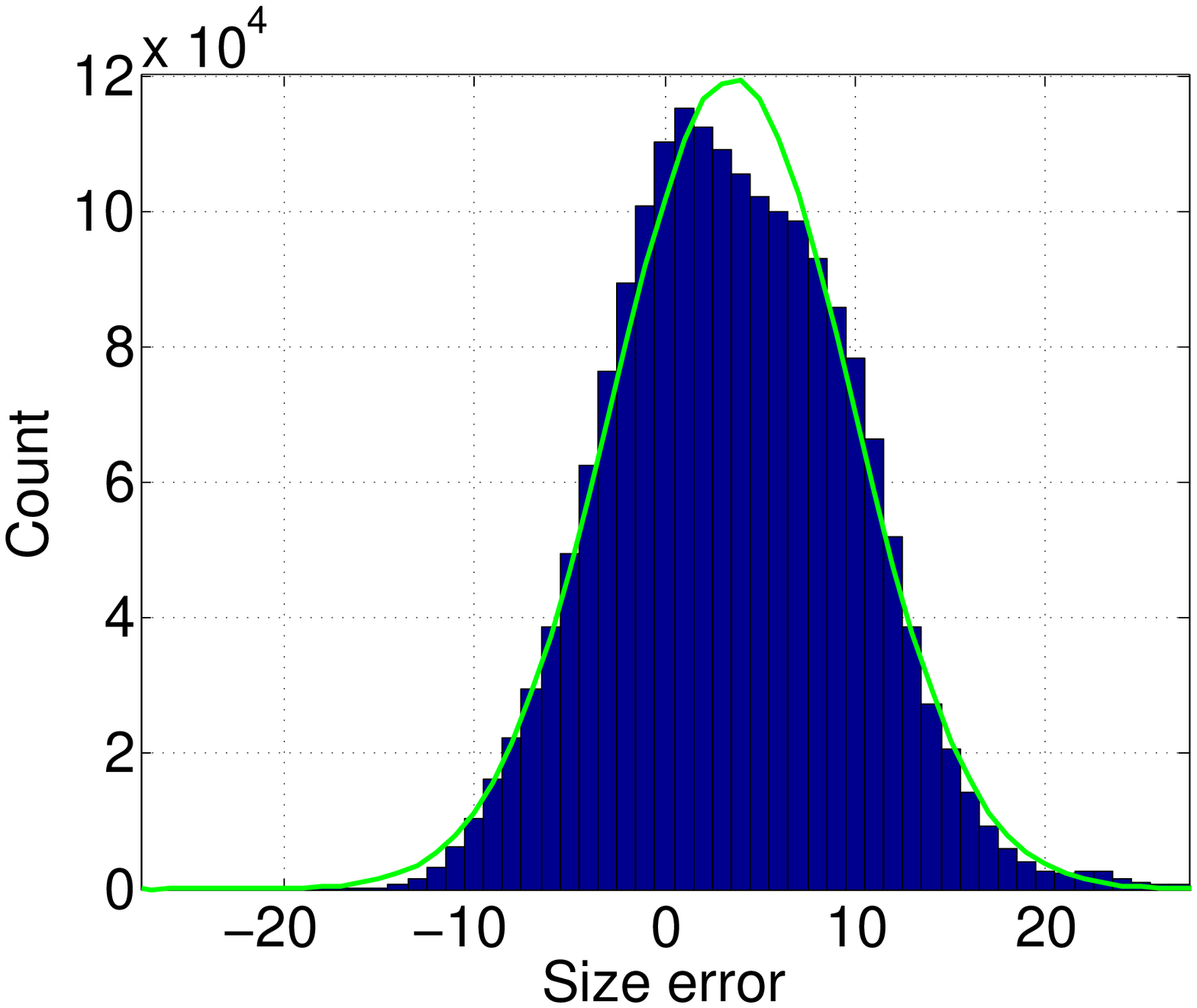}}~~
 \subfloat[X flares, $\mu_e=5.6$, $\sigma_e=6.1$.]{\plotone{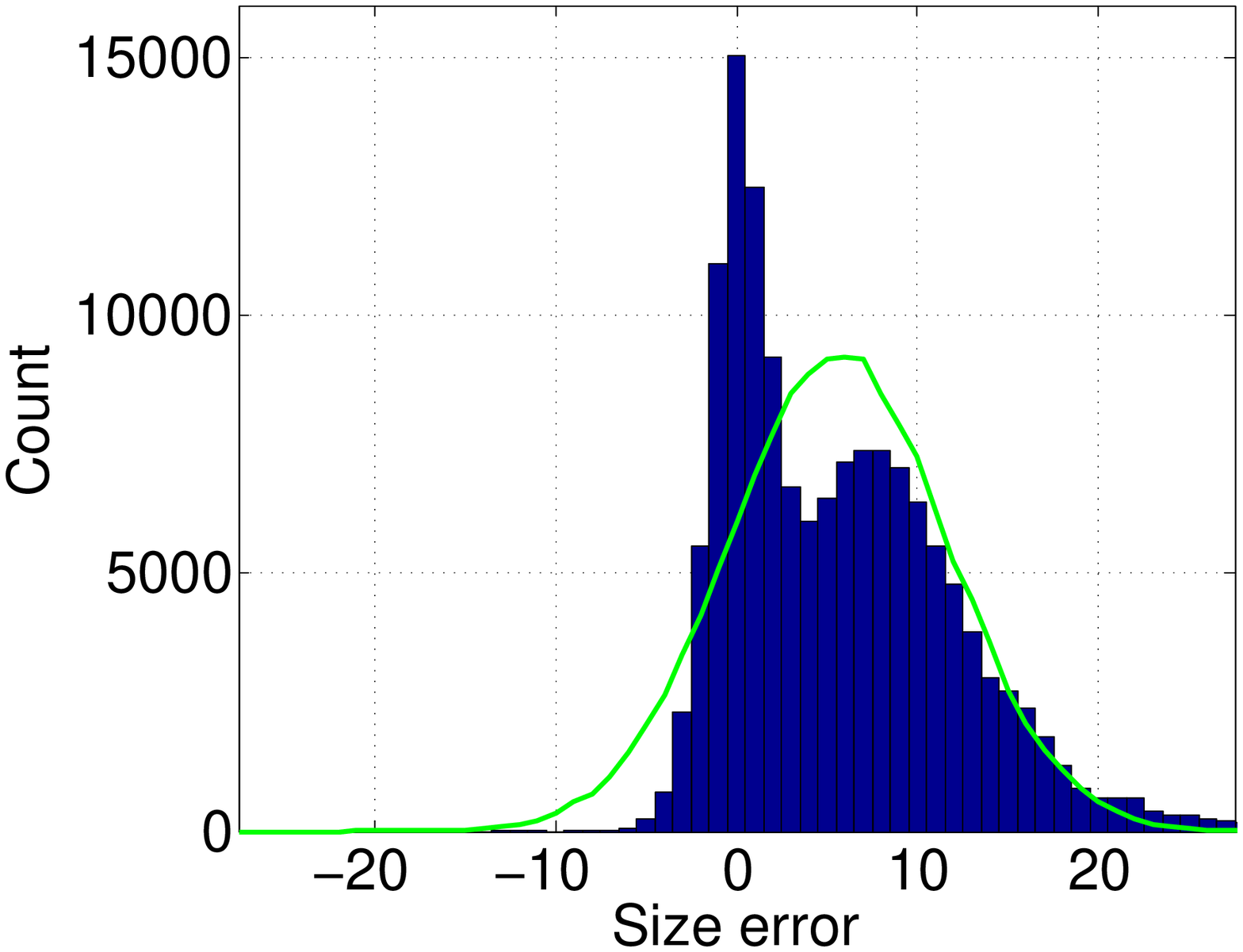}}
 \caption{Histogram of size regression error $t-\hat{t}$ for Experiment 2 (size regression, with non-flaring regions).  The sample mean $\mu_e$ and sample standard deviation $\sigma_e$ are used to plot a representative normal distribution (the green lines).}
 \label{fig:size_error_hist2}
\end{figure}

In Fig.~\ref{fig:size_error_hist2}(a) we plot a histogram of the error $t-\hat{t}$, aggregated over the 12 predictive time windows and the 100-fold cross validation runs.  Here, we see a distribution that deviates significantly from normal.  As reference, the green line in Fig.~\ref{fig:size_error_hist2}(a) is a sample normal distribution with the sample mean -3.9, sample standard deviation 5.6, and the same total counts as the error histogram. We separately plot the error histogram for non-flaring regions and the different flare classes in Fig.~\ref{fig:size_error_hist2}(b)-(f).  We see that the distribution of error for non-flaring regions (Fig.~\ref{fig:size_error_hist2}(b)) is the source of the non-normal distribution.  Furthermore, we find that most of the non-flaring regions were overestimated by the regressor, i.e., we predicted a flare when the \emph{GOES} catalog does not indicate one.  Some of this error can be attributed to the \emph{GOES} background flux as previously discussed.  Overall, we again find different average errors for different flare classes, with a tendency to overestimate small flares and underestimate large flares. 

Next, we look at the flare prediction accuracy by thresholding the regression results.  We define a non-flaring AR to be one with a predicted flare size of $\hat{t}\le\theta$ and flaring AR to be one with a predicted flare size of $\hat{t}>\theta$ where $\theta$ is the chosen threshold.  Metrics of flare prediction accuracy will be computed from the entries of the confusion matrix shown in Table~\ref{tab:confusion_matrix}. In Table~\ref{tab:confusion_matrix} TP is the true positive (correct flare prediction), FN false negative (incorrect no-flare prediction), FP false positive (incorrect flare prediction), and TN true negative (correct no-flare prediction).  Since flares are rare events, the overall prediction accuracy $(TP+TN)/(TP+FN+FP+TN)$ can be misleading.  As such, we also compute the percentage of correctly classified flaring regions (the true positive rate or TPR, also known as the sensitivity)~\citep{al-ghraibah2015}
\begin{equation}
TPR = \frac{TP}{TP+FN},
\end{equation}
the percentage of correctly classified non-flaring regions (the true negative rate or TNR, also known as the specificity)~\citep{al-ghraibah2015}
\begin{equation}
TNR = \frac{TN}{TN+FP},
\end{equation}
the Heidke skill score~\citep{bloomfield2012,al-ghraibah2015}
\begin{equation}
HSS = \frac{2[(TP\times TN)-(FN\times FP)]}{(TP+FN)(FN+TN)+(TP+FP)(FP+TN)},
\end{equation}
and the true skill score (TSS)~\citep{bloomfield2012,al-ghraibah2015}
\begin{equation}
TSS = \frac{TP}{TP+FN}-\frac{FP}{FP+TN}.
\end{equation}

\begin{figure}
 \epsscale{.40}
 \centering
 \subfloat[TPR and TNR.]{\plotone{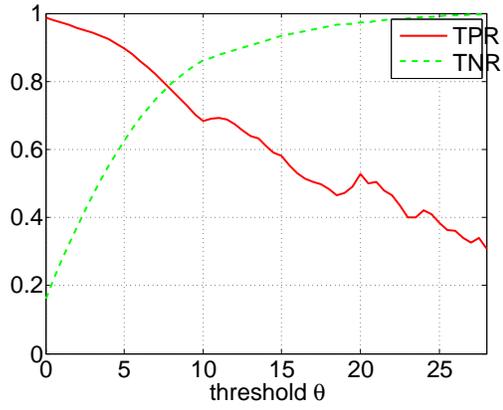}}~~
 \subfloat[TSS and HSS.]{\plotone{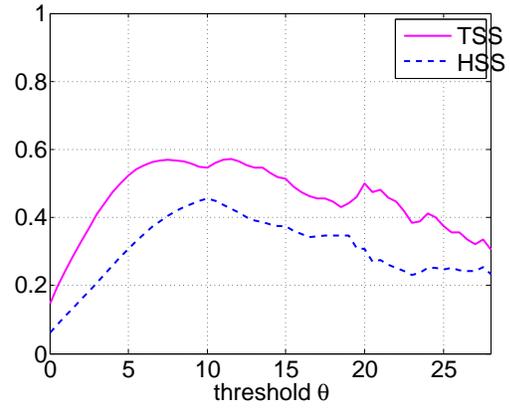}}
 \caption{Flare prediction accuracy for Experiment 2 (size regression, with non-flaring regions).}
 \label{fig:prediction_accuracy}
\end{figure}

We vary the threshold $\theta\in[0,28]$ in steps of 0.5.  Note that a threshold of $\theta=0$ corresponds to all flaring regions in this experiment, a threshold of $\theta=10$ corresponds to all flares $\ge$C1.0, a threshold of $\theta=19$ corresponds to all flares $\ge$M1.0 and a threshold of $\theta=28$ corresponds to all flares $\ge$X1.0. We plot the TPR and TNR versus $\theta$ in Fig.~\ref{fig:prediction_accuracy}(a) and the TSS and HSS versus $\theta$ in Fig.~\ref{fig:prediction_accuracy}(b).  As reference, using the same 38 features, \citet{al-ghraibah2015} obtained TPR $\approx$ 0.8, TNR $\approx$ 0.7, TSS $\approx$ 0.5, and HSS $\approx$ 0.4 for flares $\ge$C1.0.  At a comparable threshold $\theta=10$, we find performance in this work to be TPR $=0.68$, TNR $=0.86$, TSS $=0.55$, and HSS $=0.46$.  We thus find a comparable performance with a slightly lower TPR, higher TNR, and higher TSS. This is not surprising in that we are using the same 38 features as in the flare prediction work of \citet{al-ghraibah2015} and are instead usng a more general continuous-valued label vector rather than a discrete two-class label vector.

\subsection{Time Regression}
In this section, in Experiment 3, we use a label vector of the time-to-flare in hours for flaring regions.  Experiment 3 is designed to study the discriminatory potential of the 38 magnetogram complexity measures for determining time-to-flare; we note that in an operational scenario, the AR could first be classified as flaring by a separate classifier (e.g.,~\citet{al-ghraibah2015}).  

\subsubsection{Experiment 3: Time Regression, No Non-flaring Regions} 
\label{sec:experiment3}
Parameters $C$ and $\sigma$ are optimized via grid search using the same tuning dataset as Experiments 1 and 2 but no data balancing as discussed in Section~\ref{sec:experiment}.  Instead, we use 500 randomly selected data points (approximately 10\% chosen based on computational considerations) from the tuning dataset for training and the remaining for testing.  We find a minimum RMSE of approximately 39.5 hr for $C=30$ and $\sigma=0.9$ and use these values for subsequent time regression simulations.

We implement 100-fold cross validation and train over the remaining $3/4$ of the data.  For each fold, we randomly choose 1000 data points to serve as the training data and the remainder for the test data.  The choice of 1000 training points is motivated by computational considerations and by the approximate size of the balanced training set in Experiments 1 and 2.  Averaged across the 100 cross validation runs, we find an RMSE of 42.7$\pm$0.5 hr, MAE of 29.1$\pm$0.2 hr, RME of 11.5$\pm$1.5 hr, and $R$ of 0.44$\pm$0.0.  

\begin{figure}
 \epsscale{.40}
 \centering
 \plotone{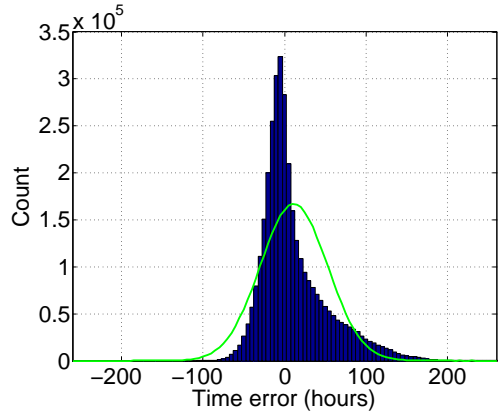}
 \caption{Histogram of time regression error $t-\hat{t}$ for Experiment 3 (time regression, no non-flaring regions).}
 \label{fig:time_error_hist}
\end{figure}

Figure~\ref{fig:time_error_hist} shows the histogram of the time prediction error. We see that the distribution is appears to be a double-sided exponential with a more heavy tail on the positive side with a sample mean of $\mu_e=11.5$ and a sample standard deviation of $\sigma_e=41.1$.  The positive value of MAE and the heavy right tail on the error distribution indicate a bias toward underestimating time to flare.  Overall, we find the time regression problem to be more error prone than the size regression problem.  This is not surprising given the consistency of the flare size regression results across the different time windows.  It appears (see discussion in Section~\ref{sec:experiment1}) that the magnetic complexity features are stable across a large window of time, indicating that these magnetic complexity features may not be particularly discriminative for time-based estimates of flaring.  

\section{Conclusion}
\label{sec:conclusions}
We have studied the prediction of solar flare size and time-to-flare using 38 features describing magnetic complexity of the photospheric magnetic field from a large set of MDI data.  The 38 magnetic complexity features were previously shown to have discriminative potential for flare prediction in~\citet{al-ghraibah2015} and were studied here for their discriminative potential in determining the size of a future flare and when a future flare will occur.  To the best of our knowledge, these regression problems have not been studied before.

This work uses support vector regression (SVR), considered state-of-the-art for time series prediction, to formulate a mapping from the 38-dimensional feature space to a continuous-valued label vector representing flare size or time-to-flare.  In the study of size prediction, we have defined a linearly spaced label vector such that an error of 1 arbitrary unit can be interpreted independently of the underlying flare class.  In the study of time-to-flare prediction, we use a label vector with the number of hours to flare.

In size regression, when we consider flaring regions only, we find an average error in estimating flare size of approximately half a \emph{GOES} class.  When we additionally consider non-flaring regions, we find an increased average error of approximately 3/4 a \emph{GOES} class. In both of these experiments, we find a different error distribution for different flare classes, with a tendency to overestimate small flares and underestimate large flares.  We also consider thresholding the regressed flare size for the experiment containing both flaring and non-flaring regions and find a true positive rate of 0.69 and a true negative rate of 0.86 for flare prediction.  These results are consistent with previous studies of flare prediction using these magnetic complexity features~\citep{al-ghraibah2015}, with a slightly lower true positive rate and slightly higher true negative rate.  The results for both of these size regression experiments are consistent across a wide range of predictive time windows, indicating that the magnetic complexity features may be persistent in appearance long before flare activity.  

This conjecture of persistence in appearance across time is supported by our larger error rates of some 40 hr in the time-to-flare regression problem.  The 38 magnetic complexity features considered here appear to have discriminative potential for flare size, but their persistence in time makes them less discriminative for the time-to-flare problem.  

Future scientific work will focus on more study of the discriminative potential of different magnetic complexity features for prediction of flare size as well as the study of the persistence of these features across time.  Algorithmically, we also plan to implement other regression methods, namely relevance vector regression, to avoid the computational overhead of optimizing kernel parameters.  Additionally, relevance vector regression will allow for a study of the underlying confidence in the regressed values.  Other methods of dealing with unbalanced data, e.g., by penalizing different errors differently~\citep{brown2000}, may alleviate the differences in error across the flare classes. Finally, the use of more recent HMI data will allow for a larger and richer dataset for training and testing these regression methods.

\clearpage
\begin{deluxetable}{lll}
 \tablewidth{0pt}
\tablecaption{Extracted features.\label{tab:features}}
\startdata
\tableline
\textbf{Gradient Features} & \textbf{Flux Evolution Features}\\
~Gradient mean                & ~FE sum\\
~Gradient std                 & ~FE absolute sum\\
~Gradient median              & ~FE gradient sum\\
~Gradient min                 & ~FE 3$\sigma$ area\\
~Gradient max                 & ~FE mean\\
~Gradient skewness            & ~FE std\\
~Gradient kurtosis            & ~FE median\\
\textbf{Neutral Line Features}   & ~FE min\\
~NL length                    & ~FE max\\
~NL no. fragments             & \textbf{Wavelet Features}\\
~NL gradient-weighted length & ~Wavelet energy level 1\\
~NL curvature mean           & ~Wavelet energy level 2\\
~NL curvature std            & ~Wavelet energy level 3\\
~NL curvature median         & ~Wavelet energy level 4\\
~NL curvature min            & ~Wavelet energy level 5\\
~NL curvature max            & \textbf{Flux Features}\\
~NL bending energy mean      & ~Total unsigned flux\\
~NL bending energy std       & ~Total signed flux\\
~NL bending energy median    & ~Total negative flux\\
~NL bending energy min       & ~Total positive flux\\
~NL bending energy max       &
\enddata
\end{deluxetable}

\clearpage
\begin{deluxetable}{lll}
 \tablewidth{0pt}
\tablecaption{Flare sizes in W m${}^{-2}$ and regression target values.\label{tab:flare_sizes}}
\startdata
\tableline
Class & Peak Flux (W m${}^{-2}$)  & Regression Target Value\\
\tableline
A     & $<10^{-7}$           & 0.1--1\\
B     & $10^{-7}$--$10^{-6}$ & 1--9\\
C     & $10^{-6}$--$10^{-5}$ & 10--18\\
M     & $10^{-5}$--$10^{-4}$ & 19--27\\
X     & $>10^{-4}$           & 28--36+
\enddata
\end{deluxetable}

\clearpage
\begin{deluxetable}{lcc}
 \tablewidth{0pt}
\tablecaption{Confusion matrix for flare prediction by thresholding the predicted flare size.\label{tab:confusion_matrix}}
\startdata
\tableline
& \multicolumn{2}{c}{Forecasted}\\
Observed & Flare & No Flare\\\tableline
Flare & TP & FN\\
No Flare & FP & TN
\enddata
\end{deluxetable}

\clearpage

\clearpage

\clearpage

\clearpage

\clearpage

\clearpage






\clearpage

\end{document}